\documentstyle[epsfig,12pt]{article}

\def\beq{\begin{equation}}
\def\eeq{\end{equation}}

\def\be{\begin{equation}}
\def\bea{\begin{eqnarray}}
\def\ee{\end{equation}}
\def\eea{\end{eqnarray}}
\def\d{\partial}
\def\eqref#1{(\ref{#1})}

\def\a{\alpha}

\def\bra{\langle}
\def\ket{\rangle}
\def\e{{\rm e}}
\def\tr{{\rm tr}}

\begin{document}


\begin{titlepage}

\begin{centering}

\vspace*{3cm}

{\Large\bf ${\cal N}$=1 super Yang--Mills on a (3+1) dimensional
transverse lattice with one exact supersymmetry}

\vspace*{1.5cm}

{\bf  Motomichi Harada and Stephen Pinsky}
\vspace*{0.5cm}

{\sl Department of Physics \\
Ohio State University\\
Columbus OH 43210}

\vspace*{0.5cm}

\vspace*{1cm}


\vspace*{1cm}

\vspace*{1cm}

\begin{abstract}
We formulate ${\cal N}$=1 super Yang-Mills theory in 3+1 dimensions
on a two dimensional transverse lattice using supersymmetric
discrete light cone quantization in the large-$N_c$ limit. This
formulation is free of fermion species doubling. We are able to
preserve one supersymmetry. We find a rich, non-trivial behavior of
the mass spectrum as a function of the coupling $g\sqrt{N_c}$, and
see some sort of ``transition" in the structure of a bound state as
we go from the weak coupling to the strong coupling. Using a toy
model we give an interpretation of the rich behavior of the mass
spectrum. We present the mass spectrum as a function of the winding
number for those states whose color flux winds all the way around in
one of the transverse directions. We use two fits to the mass
spectrum and the one that has a string theory justification appears
preferable. For those states whose color flux is localized we
present an extrapolated value for $m^2$ for some low energy bound
states in the limit where the numerical resolution goes to infinity.
\end{abstract}
\end{centering}

\vfill

\end{titlepage}
\newpage

\section{Introduction}

In the past years, there has been a tremendous amount of progress in
the analytical understanding of supersymmetric theories due to the
discovery of AdS/CFT correspondence \cite{Maldacena:1997re}, using
``orbifolding" \cite{Kachru:1998ys}, ``orientifolding"
\cite{Armoni:2003gp,Armoni:2003fb}, or others. Recently Armoni,
Shifman and Veneziano have shown that in the large $N_c$ limit a
non-supersymmetric gauge theory with a Dirac fermion in the
antisymmetric tensor representation is equivalent, both
perturbatively and nonperturbatively, to ${\cal N}$=1 super
Yang-Mills (SYM) theory in its bosonic sector \cite{Armoni:2003gp}.
Since for $N_c=3$ the non-supersymmetric gauge theory is just
one-flavor QCD, even though we have to keep in mind $1/N_c$
corrections, knowing the nonperturbative properties of ${\cal N}$=1
SYM is of great importance from phenomenological viewpoint as well.
Accordingly, one cannot stress enough the importance of being able
to perform some nonperturbative numerical calculations for ${\cal
N}$=1 SYM to test the predictions made by Armoni, Shifman and
Veneziano. However, this is not an easy task by any means. This is
because the supersymmetry (SUSY) transformation, which is an
extension of the Poincar\'{e} transformation, is broken on a lattice
due to the lack of continuous spatial translational symmetry and
because Leibniz rule does not hold on a lattice. Therefore, new
attempts to put SYM on the lattice are interesting and eagerly
awaited.

There have been some promising approaches along this direction
\cite{Cohen:2003xe, Sugino:2003yb, Sugino:2004gz} that partially
preserve SUSY. However, there still appear to be some remaining
issues before the method introduced by Cohen, Kaplan, Katz and Unsal
becomes a practical computational approach \cite{Cohen:2003xe,
Giedt:2003ve}. The approach by Sugino \cite{Sugino:2003yb,
Sugino:2004gz} has encountered unwanted surplus modes in four
dimensions (in Euclidean space) \cite{Sugino:2004gz} and is still
waiting numerical simulations\footnote{We should note that the
topological field approach to constructing a supersymmetric theory
on a lattice utilized by Sugino was first discussed by Catterall in
Ref.\cite{Catterall:2003wd}, which investigates theories without a
gauge symmetry. Very recently, Catterall has proposed a geometrical
approach to ${\cal N}$=2 SYM on the two dimensional lattice in
Ref.~\cite{Catterall:2004np}.}. For other recent progress in an
effort to realize SUSY on a lattice, see for example Ref.
\cite{Bietenholz:1998qq, Feo:2002yi}.

Recently the authors proposed another method to put SYM on a lattice
\cite{Harada:2003bs}. This approach is in fact not a new idea,
rather it is motivated by the idea of the ``(de)construction"
\cite{Hill:2000mu} and is a mixture of the two existing ideas; the
transverse lattice \cite{bardeen,rev} and supersymmetric discrete
light cone quantization (SDLCQ) \cite{Matsumura:1995kw}. Our first
attempt was made for 2+1 dimensional ${\cal N}=1$ SYM with one
transverse lattice. Here we present a formulation for 3+1
dimensional ${\cal N}$=1 SYM with a two dimensional transverse
lattice in the large $N_c$ limit.

At each site of the two dimensional lattice, we have one gauge boson
and one four-component Majorana spinor. Adjacent sites are connected
by the link variables. All these fields depend only on the
light-cone time and spatial coordinates $x^{\pm}$ and are associated
with two site indices, say $(i,j)$. In the large $N_c$ limit,
however, it turns out that we are allowed to drop the site indices
for our calculation. This is in some sense the manifestation of the
Eguchi-Kawai reduction \cite{egk83}. However, it is well known that
the naive Eguchi-Kawai reduction encounters a problem due to the
violation of one of the assumptions made by Eguchi and Kawai
\cite{Bhanot:1982sh}. That assumption is the $U(1)^d$ symmetry.
Since we do not have to assume the $U(1)^d$ symmetry to justify our
reduction of the transverse lattice degrees of freedom, we believe
that we do not have to introduce quenching \cite{Bhanot:1982sh} or
twisted \cite{Gonzalez-Arroyo:1982hz} lattices, which were invented
to overcome the problem associated with the naive Eguchi-Kawai
reduction at weak couplings. For more complete and detailed
discussion for this claim, see appendix \ref{reduction}. With this
reduction of the transverse degrees of freedom, we can regard all
the fields as 1+1 dimensional objects. That is to say that we have
some complicated 1+1 dimensional field theory with some highly
non-trivial interactions of the fields. Furthermore, since we can
always work in the frame where we have zero transverse momenta
$P^1,P^2=0$, ${\cal N}$=1 SUSY algebra in 3+1 dimensions becomes
identical to ${\cal N}$=2 SUSY algebra in 1+1 dimensions, which is
sometimes referred to as ${\cal N}$=(2,2) SUSY in literature, (2,2)
for two $Q^+$'s and two $Q^-$'s. We are able to maintain one of this
underlying ${\cal} N$=(2,2) SUSY algebra in our formulation, meaning
that we are able to preserve one exact SUSY.

We discretize light-cone momentum $p^+$ by imposing the periodic condition
on the light-cone spatial coordinate $x^-$. Thus, we have two spatial
lattices and one momentum lattice in our model. Since we are dealing with
spatial lattices, one has to be concerned about the notorious fermion
doubling problem. In fact it is well known that the transverse lattice
suffers from the doubling problem \cite{Burkardt:1998ws}. However,
the authors have found that SDLCQ formulation of a transverse lattice model
is automatically free of the doubling problem \cite{Harada:2004cc}.

There are some aspects of this calculation that are similar to the
2+1 dimensional model \cite{Harada:2003bs} and there are others that
are not. What is not the same is that the supercharge $Q^-_{\a}$ has
terms which have different powers of the coupling $g'\equiv
g\sqrt{N_c}$, where $\a=1,2$. To be more precise, $Q^-_{\a}$
consists of terms proportional to $g'$ and terms proportional to
$g'^3$. The different powers of $g'$ give rise to a rich spectrum as
one varies $g'$, and the wavefunctions depend on $g'$. This means
that it is possible to see wavefunctions which are almost vanishing
at small couplings, but become very large at strong couplings, and
vice versa.

One more thing which is different from the previous case is that our
$Q^-_{\a}$ has terms of third and fifth order in dynamical fields, while
all of the terms in $Q^-$ are of third order for 2+1 dimensional
case. This leads to a hamiltonian of eighth order in fields, which
is of higher order than the hamiltonian of sixth order that we get
from the standard formulation of 3+1 dimensional ${\cal N}$=1 SYM on
the two transverse lattice. We admit that this is a disadvantage of
our formulation in 3+1 dimensions compared to that in 2+1
dimensions. Nevertheless, we still think that our approach is more
advantageous since in the SDLCQ formulation we use $Q^-_{\a}$, not the
hamiltonian, and this $Q^-_{\a}$ is still of lower order in fields than
the hamiltonian obtained from the standard formulation, and since
the standard formulation suffers from the fermion doubling problem.

Similar to the 2+1 dimensional case we are not able to preserve the
full supersymmetry algebra. We are able to maintain one exact SUSY.
This is attributed to the fact that when quantizing the dynamical
fields we have to make the link variable, which is a unitary matrix,
a linear complex matrix. One way to compensate for the effects of
this ``linearization" is to make use of the ``color-dielectric"
formulation of the lattice gauge theory
\cite{rev,Dalley:1998bj,Dalley:1997an}. In this formulation we
consider smeared degrees of freedom ${\cal M}$, which are obtained
from the original link variable $M$ by averaging $M$ over some
finite volume, say $\sum_{av}M$. In order for this smeared theory to
be equivalent to the original one, we must have an effective
potential for the ${\cal M}$ defined by integrating out $M$
\cite{Dalley:1998bj}
\[ \exp[-V_{eff}({\cal M})]=\int {\cal D}M\delta({\cal
M}-\sum_{av}M)\exp[-S_{canonical}(M)]. \]%
However, this $V_{eff}({\cal M})$ can be very complicated and
performing the path integral above is extremely difficult, if not
impossible. Thus, one makes some approximations with ansatz to
determine $V_{eff}$. For more detail, we'd refer the reader to the
Refs. \cite{rev,Dalley:1998bj,Dalley:1997an}.

To constrain the linearized fields, we require the model to exactly
conserve one SUSY as we did for our 2+1 dimensional calculation.
That is, we present a physical $Q^-_{\a}$ that preserves one SUSY. By
``physical" we mean a $Q^-_{\a}$ which transforms one physical state into
another physical state. We are not able to fully recover SUSY due to
the absence of a physical $Q^+_{\a}$. This defect results in a different
number of massless states in the bosonic and fermionic sectors.
However, we do see the mass degeneracy among the massive bosonic and
fermionic states. The linearization doubles the bosonic degrees of
freedom, leading to the SUSY breakdown. The partial recovery of SUSY
implies that we have cured some but not all of the problems
associated with the linearization.

We are numerically able to identify what we call the cyclic states
and non-cyclic states by examining the properties of the states. The
cyclic states are those whose color flux winds all the way around in
one or two of the transverse directions. For the non-cyclic states
the color flux is localized in color space. The cyclic bound states
have a non-trivial spectrum as a function of the winding number. We
find that $m^2$ for the cyclic bound states can be fit by either
$b+c/W_I+d/W_I^2$ or $b+cW_I^2+d/W_I^2$, where $b,c,d$ are some
constants and $W_I$ is the winding number in the $x_I$-direction
with $I=1,2$. It could be interesting to know how the form of the
$m^2$ changes from weak coupling to strong coupling however the
complicated spectrum for strong couplings puts this beyond our reach
at the present time.

The structure of this paper is the following. In Sec.~2 we present a
standard formulation of ${\cal N}$=1 SYM with a two dimensional
transverse lattice and derive constraint equations on the physical
states. We discuss the implications of those in some detail. We give
SDLCQ formulation of ${\cal N}$=1 SYM in Sec.~3 and show that this
formulation is free from the doubling problem. The coupling
dependence of the mass spectrum is discussed in Sec.~4 followed by
numerical results for cyclic bound states in Sec.~5 and for
non-cyclic bound states in Sec.~6. The summary and possible further
directions of investigation are given in Sec.~7. Appendix
\ref{reduction} is to show how we justify the reduction of
transverse degrees of freedom in the large $N_c$ limit. Derived in
the appendix \ref{majorana} is the ${\cal N}$=1 SUSY algebra in
Majorana representation in the $D+1$ dimensional light-cone
coordinates with $D=1,2,3$.

\section{Transverse lattice model in 3+1 dimensions}
\label{sec:translat}
In this section we present the standard formulation of a transverse
lattice model in 3+1 dimensions for an ${\cal N}$=1 supersymmetric
$SU(N_c)$ theory with adjoint bosons and adjoint fermions in the
large--$N_c$ limit. We work in light-cone coordinates so that
$x^{\pm}\equiv (x^0\pm x^3)/\sqrt 2$. The metric is specified by
$x^\pm=x_\mp$ and $x^{I}=-x_{I}$, where ${I=1,2}$. Suppose that
there are $N_{sites}$ sites in both the transverse directions $x^1$
and $x^2$ with lattice spacing $a$. With each site, say $n=(i,j)$,
we associate one gauge boson field $A_{\nu,n}(x^{\mu})$ and one
four-component Majorana spinor $\Psi_n(x^{\mu})$, where $\nu,
\mu=\pm$. $A_{\nu,n}$'s and $\Psi_n$'s are in the adjoint
representation. The adjacent sites, say $n$ and $n+i_{I}$, where
$i_{I}$ is a vector of length $a$ in the direction $x^{I}$, are
connected by what we call the link variables $M_n^{I}(x^{\mu})$ and
$M_n^{I\dag}(x^{\mu})$. $M_n^{I}(x^{\mu})$ stands for a link which
goes from the site $n$ to the site $(n+i_{I})$ and
$M_n^{I\dag}(x^{\mu})$ for a link from the site $(n+i_{I})$ to $n$.
We impose the periodic condition on the transverse sites so that
$A_{N_{sites}i_{I}+n}=A_n$, $\Psi_{N_{sites}i_{I}+n}=\Psi_n$,
$M^{I}_{N_{sites}i_{I}+n}=M_n^{I}$ and
$M^{I\dag}_{N_{sites}i_{I}+n}=M^{I\dag}_n$. Under the transverse
gauge transformation \cite{rev} the fields transform as
\begin{equation}
gA_n^{\mu} \longrightarrow
U_ngA_n^{\mu}U_n^{\dag}-iU_n\partial^{\mu}
  U_n^{\dag}, \quad
  M_n^{I}\longrightarrow U_nM_n^{I}U_{n+i_{I}}^{\dag}, \quad
  \Psi_n \longrightarrow U_n\Psi_nU_n^{\dag},
\label{gauge}
\end{equation}
where $g$ is the coupling constant and $U_n \equiv U_n(x^{\mu})$
is a $N_c \times N_c$ unitary matrix. In all earlier work on the
transverse lattice \cite{rev} $\Psi_n$ was in the fundamental
representation.

The link variable can be written as
\begin{equation}
  M_n^{I}(x^{\mu})=\exp\left( iagA_{n+i_{I}/2,I}(x^{\mu})\right), \label{M}
\end{equation}
where $A_{n,I}$ is the transverse component of the gauge potential
at site $n$ and as $a \to 0$ we can formally expand Eq. \eqref{M} in
powers of $a$ as follows:
\begin{equation}
M_n^{I}(x^{\mu})
   =1+iagA_{n,I}(x^{\mu})+\frac {a^2}2 \left[ ig \partial_{I}
   A_{n,I}(x^{\mu})-g^2 \left(A_{n,I}
   (x^{\mu})\right)^2 \right]+O(a^3).
\label{expandM}
\end{equation}
 In the limit $a \to 0$, with the substitution of the expansion Eq.~\eqref{expandM} for
$M_n^{I}$, we expect everything to coincide with its counterpart in
{\it continuum} (3+1)--dimensional theory.

The discrete Lagrangian is then given by
\begin{eqnarray*}
&&{\cal L}
 =\tr\Bigg\{-\frac 14 F_{n}^{\mu\nu}F_{n,\mu\nu}+\frac 1{2a^2g^2}
   (D_{\mu}M_{n}^{I})(D^{\mu}M_{n}^{I})^{\dag} \\
&&+\frac 1{4a^4g^2}\sum_{I\ne J}( M_{n}^{I}
   M_{n+i_{I}}^{J}M_{n+i_{J}}^{I\dag}
   M_{n}^{J\dag}-1)
   +\bar{\Psi}_{n}i\Gamma^{\mu}D_{\mu}\Psi_{n} \\
&&+\frac i{2a}\bar{\Psi}_{n}\Gamma^{I}(M_{n}^{I}\Psi_{n+i_{I}}
M_{n}^{I\dag}-M_{n-i_{I}}^{I\dag}\Psi_{n-i_{I}}M_{n-i_{I}}^{I})\Bigg\},
\end{eqnarray*}
where the trace has been taken with respect to the color indices,
$F_{n,\mu\nu}=
\partial_{\mu}A_{n,\nu}-\partial_{\nu}A_{n,\mu}+ig[A_{n,\mu},A_{n,\nu}]$, $\mu$, $\nu$ =
$\pm$. We choose Majorana representation where Majorana spinors have
real component fields and $\Gamma$'s are given by
\[ \Gamma^0\equiv \left(\begin{array}{cc} 0&\sigma_2 \\ \sigma_2&0\end{array}
   \right), \quad
   \Gamma^1\equiv i\left(\begin{array}{cc} \sigma_1&0 \\ 0&\sigma_1\end{array}
   \right),\quad
   \Gamma^2\equiv i\left(\begin{array}{cc} \sigma_3&0 \\ 0&\sigma_3\end{array}
   \right),\quad
   \Gamma^3\equiv \left(\begin{array}{cc} 0&-\sigma_2 \\ \sigma_2&0\end{array}
   \right),
\]
\[  \Gamma^{+}\equiv\frac {\Gamma^0+\Gamma^3}{\sqrt 2}=\left(\begin{array}{cc}
    0&0 \\\sqrt 2\sigma_2& 0\end{array}\right),\quad
    \Gamma^{-}\equiv\frac {\Gamma^0-\Gamma^3}{\sqrt 2}=\left(\begin{array}{cc}
    0&\sqrt 2\sigma_2 \\ 0& 0\end{array}\right).
\]
The covariant derivative $D_{\mu}$ is defined by
\begin{eqnarray*}
&&D_{\mu}\Psi_{n}\equiv \d_{\mu}\Psi_{n}+ig[A_{n,\mu},\Psi_{n}],
\\ &&D_{\mu}M_{n}^{I}\equiv
\d_{\mu}M_{n}^{I}+igA_{n,I}M_{n}^{I}
  -igM_{n}^{I}A_{n+i_{I},\mu}
  \stackrel{a \to 0}{\longrightarrow} iag F_{\mu I} +O(a^2), \\
&&(D^{\mu}M_{n}^{I})^{\dag}\equiv \d^{\mu}M_{n}^{I\dag}
  -igM_{n}^{I\dag}A_{n}^{\mu}+igA_{n+i_{I}}^{\mu}
  M_{n}^{I\dag}
  \stackrel{a \to 0}{\longrightarrow} iag F^{\mu I} +O(a^2).
\end{eqnarray*}
In the limit $a\to 0$ we recover the standard Lagrangian as
expected. Of course the form of this Lagrangian is slightly
different from that in Ref. \cite{rev} since the fermions are in
the adjoint representation. This  Lagrangian is hermitian and
invariant under the transformation in Eq.~\eqref{gauge} as one
would expect.

The following Euler-Lagrange equations in the light cone gauge,
$A_{n,-}=0$, are constraint equations.
\begin{eqnarray}
&&  \partial_-^2A_n^-\equiv gJ_n^+
   \stackrel{a \to 0}{\longrightarrow} ig[A_{I},\d_-A_{I}]+
   \d_{I}\d_-A_{I}+2g\psi_R\psi_R,    \label{equations} \\
&&  \partial_- \psi_{Ln}=\frac {-i}{2\sqrt 2a}\sigma_2\beta_{I}
  (M_n^{I}\psi_{Rn+i_{I}} M_n^{I\dag}
  -M_{n-i_{I}}^{I\dag}\psi_{Rn-i_{I}}M_{n-i_{I}}^{I}) \
   \stackrel{a \to 0}{\longrightarrow} \ \frac {-i}{\sqrt 2}
   \sigma_2\beta_{I}D_{I} \psi_R,
   \nonumber
\end{eqnarray}
where
\begin{equation}
J_n^+ \equiv
  \frac i{2g^2a^2} (M_n^{I}\stackrel{\leftrightarrow}{\partial}_-
  M_n^{I\dag}+M_{n-i_{I}}^{I\dag}\stackrel{\leftrightarrow}{\partial}_-
  M_{n-i_{I}}^{I})+2\psi_n\psi_n, \quad
  \Psi_n \equiv \frac {1}{2^{1/4}}
\left(\begin{array}{c}\psi_{Rn} \\ \psi_{Ln} \end{array}\right),
\end{equation}
$\beta_1\equiv \sigma_1$, $\beta_2=\sigma_3$ and $\psi_{L,R}$ are
the two-component left-moving, right-moving spinors.

Since these equations only involve the spatial derivative we can
solve them for $A_n^-$ and $\psi_{Ln}$, respectively. Thus the
dynamical field degrees of freedom are $M_n^{I}$, $M_n^{I\dag}$ and
$\psi_{Rn}$.

Eq. \eqref{equations} gives a constraint on physical states $|phys
\ket$, since the zero mode of $J^+_n$ acting on any physical state
must vanish,
\begin{equation}
  \stackrel{0}{J^+_n}|phys \ket=\int dx^- J^+_n(x^{\mu})|phys \ket=0 \quad
  {\rm for} \ {\rm any} \ n=(i,j).
\label{constraint}
\end{equation}
This means that the physical states must be color singlet at {\it
each} site.

It is straightforward to derive $P^{\pm}\equiv \int dx^-
T^{+\pm}$, where $T^{\mu\nu}$ is the stress-energy tensor. We have
\begin{eqnarray}
   P^+&=& a^2\sum_{n}\int dx^- \tr\left(
         \frac 1{a^2g^2}\d_-M_n^{I\dag}\d_-M_n^{I}
         +i\psi_{Rn}\d_-\psi_{Rn}\right),\label{pplus} \\
   P^-
    &=&a^2\sum_{n}\int dx^- \tr\Biggl[ \frac 12(\d_-A_n^-)^2
        +i\psi_{Ln}\d_-\psi_{Ln}\nonumber \\
       &&-\frac 1{4a^2g^2}( M_{n}^{I}
       M_{n+i_{I}}^{J}M_{n+i_{J}}^{I\dag}
       M_{n}^{J\dag}-1)\Biggr] \label{pminus},
\end{eqnarray}
where one should notice that we've kept the non-dynamical fields in
the expression for $P^-$ to make it look simpler. When one quantizes
the dynamical fields, unitarity of $M_n^{I}$ is lost and $M_n^{I}$
becomes an $N_c \times N_c$ complex matrix \cite{rev}. One way to
compensate for the effects of this ``linearization" is to make use
of the ``color dielectric" formulation of the lattice gauge theory
\cite{rev,Dalley:1998bj,Dalley:1997an}. We will approach this issue
using supersymmetry as we've done for the 2+1 dimensional case.

Having linearized $M_n^{I}$, we can expand $M_n^{I}$ and $\psi_{Rn}$
in their Fourier modes as follows; at $x^+=0$
\begin{eqnarray}
 M_{n,rs}^{I}(x^-)&=&\frac {ag}{\sqrt{2\pi}}\int_0^{\infty}
   \frac{dk^+}{\sqrt{k^+}}(d_{n,rs}^{I}(k^+)\e^{-ik^+x^-}
   +a_{n,sr}^{I\dag}(k^+)\e^{ik^+x^-}), \label{m} \\
  u_{n,rs}^{\a}(x^-)&=&\frac 1{2\sqrt{\pi}}\int_0^{\infty}  dk^+
   (b_{n,rs}^{\a}(k^+)\e^{-ik^+x^-}+b_{n,sr}^{\a\dag}(k^+)
   \e^{ik^+x^-}), \label{psi}
\end{eqnarray}
where $r,s$ indicate the color indices, $\psi_{Rn}\equiv
\left(\begin{array}{c} u^1_n \\ u^2_n \end{array}\right) $,
$\a=1,2$, $a_{n,sr}^{\dag}(k^+)$ creates a link variable with
momentum $k^+$ which carries color $r$ at site $n$ to $s$ at site
$(n+i_{I})$, $d^{\dag}_{n,sr}(k^+)$ creates a link with $k^+$ which
carries color $r$ at site $(n+i_{I})$ to $s$ at site $i$ and
$b^{\a\dag}_{n,sr}$ creates a fermion at the site $n$ which carries
color $r$ to $s$. Quantizing at $x^+=0$ we have

\begin{equation}
[M_{ij,rs}^{I}(x^-),\pi_{Mkl,pq}^J(y^-)] =
\Bigg[M_{ij,rs}^{I}(x^-),\frac
{\partial_{-y}M^{J\dag}_{kl,pq}(y^-)}{2a^2g^2}\Bigg]= \frac i2
\delta(x^--y^-)   \frac {\delta_{ik}}a \frac {\delta_{jl}}a
   \delta_{rp}\delta_{sq}\delta_{IJ},
\end{equation}
\begin{equation}
    \{ u_{ij,rs}^{\a}(x^-),\pi_{\psi kl,pq}^{\beta}(y^-)\}=
   \{ u_{ij,rs}^{\a}(x^-),iu_{kl,pq}^{\beta}(y^-)\}=\frac i2 \delta(x^--y^-)
   \frac {\delta_{ik}}a \frac {\delta_{jl}}a
   \delta_{rp}\delta_{sq}\delta_{\a\beta},
\end{equation}
where $\pi_M,\pi_{\psi}$ are the conjugate momentum for $M,\psi$,
respectively, and we wrote out the site indices for clarity. Note
that we divided $\delta_{ik}$ and $\delta_{jl}$ by $a$ because
$\delta_{ik}/a \to \delta(x^{1}-y^{1})$ and $\delta_{jl}/a \to
\delta(x^{2}-y^{2})$ as $a \to 0$. Then, one can easily see that
these commutation relations are satisfied when $a$'s, $d$'s and
$b$'s satisfy the following:

\begin{equation}
 [a_{ij,rs}^{I}(k^+),a^{J\dag}_{kl,pq}(p^+)]
 =[d_{ij,rs}^{I}(k^+),d^{J\dag}_{kl,pq}(p^+)] \nonumber \\
 =\delta(k^+-p^+)\frac {\delta_{ik}}a\frac {\delta_{jl}}a
  \delta_{rp}\delta_{sq}\delta_{IJ}, \label{commutations1}
\end{equation}
\begin{equation}
\{ b_{ij,rs}^{\a}(k^+),b^{\beta\dag}_{kl,pq}(p^+)\}
 =\delta(k^+-p^+)\frac {\delta_{ik}}a\frac {\delta_{jl}}a
  \delta_{rp}\delta_{sq}\delta_{\a\beta}, \label{commutations2}
\end{equation}
with others all being zero. Physical states can be generated by
acting on the Fock vacuum $|0 \ket$ with these $a^{I\dag}$'s,
$d^{I\dag}$'s and $b^{\a\dag}$'s in such a manner that the
constraint Eq.~\eqref{constraint} is satisfied.

Before discussing the physical constraint in more detail, let us
point out the fact that this naive Lagrangian formulation is {\it
not free from the fermion species doubling problem}, while our SDLCQ
formulation that we will introduce in the next section actually {\it
is} \cite{Harada:2004cc}. Nonetheless, the constraint equation would
still be valid since the constraint equation (\ref{equations},
\ref{constraint}) was derived from $\frac{\delta {\cal L}}{\delta
A^-_n}-\d_+\frac{\delta {\cal L}}{\delta \d_+A^-_n}=0$ in which we
do not have any problematic terms responsible for the doubling
problem, i.e. the terms which contains the difference between
fermions at different sites. Therefore, we assume that this physical
constraint is valid for our SDLCQ formulation in the next section
and we will fully utilize it when we carry out our numerical
calculations.

\begin{figure}[ht]
\begin{tabular}{cc}
\epsfig{figure=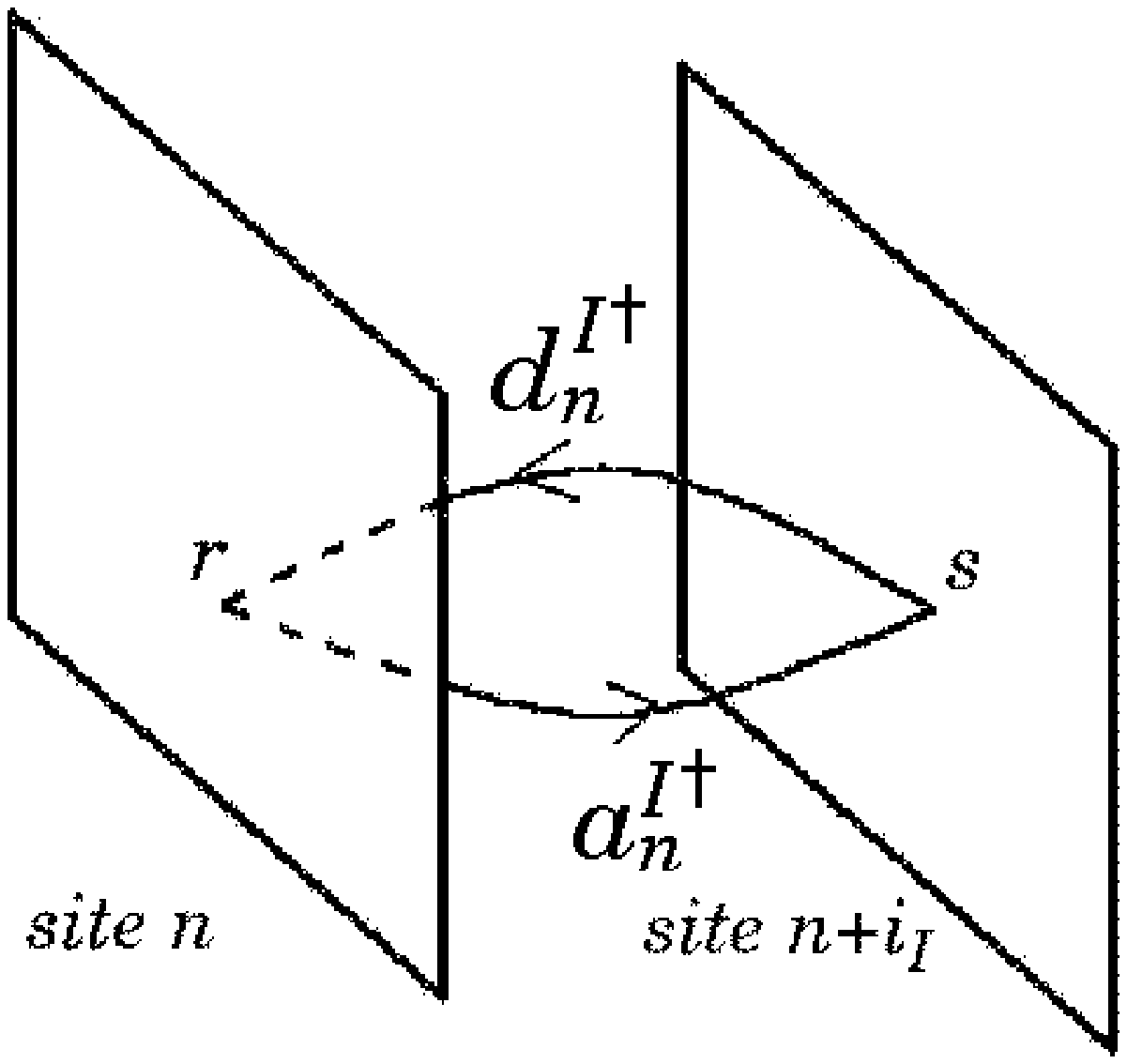,width=7.3cm,angle=0}&
\epsfig{figure=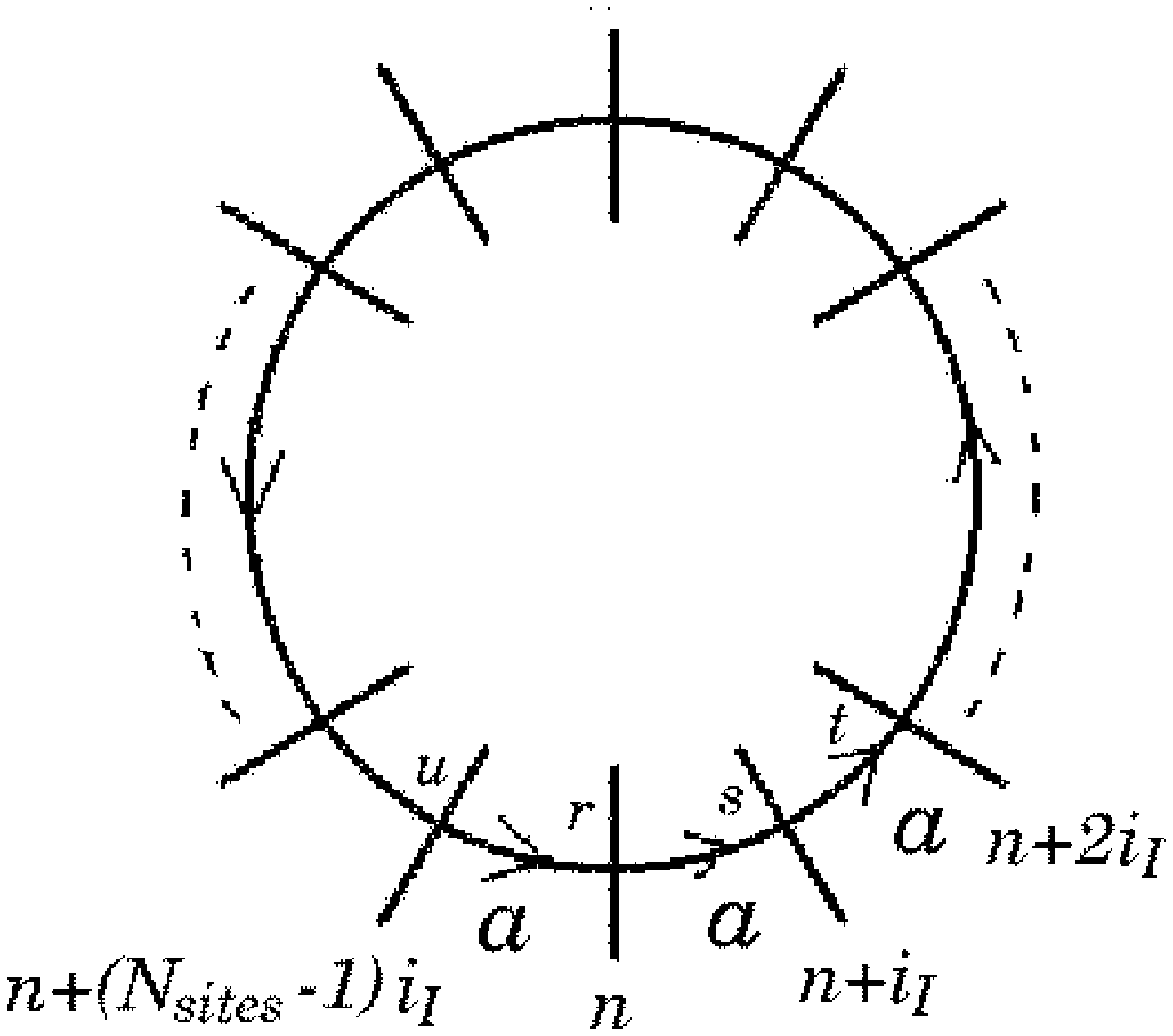,width=7.3cm,angle=0} \\
(a) & (b)
\end{tabular}

\caption{(a)The color charge for the state $|phys \ 1\ket \equiv
d^{I\dag}_{n,rs}a^{I\dag}_{n,sr}|0\ket$. The planes represent the
color space. $a_n^{I\dag}$ carries color $r$ at site $n$ to $s$ at
site $n+i_I$ and $d_n^{I\dag}$ carries it back to $r$ at site $n$.
(b) The color charge for the state $|phys \ 2\ket \equiv
a^{I\dag}_{n+(N_{sites}-1)i_I,ru}\cdots a^{I\dag}_{n+i_I,ts}
a^{I\dag}_{n,sr}|0\ket$. The lines which intersect a circle
represent the color planes at sites. The color goes all the way
around the transverse lattice.} \label{da}
\end{figure}

With this subtlety in mind, let us complete this section by
discussing the physical constraint \eqref{constraint} in more
detail. The states are all constructed in the large--$N_c$ limit,
and therefore we need only consider single--trace states.
 In order for a state to be color
singlet at each site, {\it each color index has to be contracted at
the same site}. As an example consider a state represented by $|phys
\ 1\ket \equiv d^{I\dag}_{n,rs}a^{I\dag}_{n,sr}|0\ket$, where we've
suppressed the momentum carried by $a^{I\dag}$ and $d^{I\dag}$ and
we'll do so hereafter unless it's necessary for clarity.  For this
state the color $r$ at site $n$ is carried by $a^{I\dag}_n$ to $s$
at site $(n+i_I)$ and then brought back by $d^{I\dag}_n$ to $r$ at
site $n$. The color $r$ is contracted at site $n$ {\it only} and the
color $s$ at site $(n+i_I)$ {\it only}. Therefore, this is a
physical state satisfying Eq. \eqref{constraint}. A picture to
visualize this case is shown in Fig. \ref{da}a. Diagrammatically,
one can say that at every point in color space one has to have
either no lines or two lines, one of which goes into and the other
of which comes out of the point, so that the color indices are
contracted at the same site.

One also needs to be careful with operator ordering. One can show
that the state $
d^{I\dag}_{n,rs}a^{I\dag}_{n,st}b^{I\dag}_{n,tr}|0\ket$ is physical,
while the state
$b^{I\dag}_{n,rs}a^{I\dag}_{n,st}d^{I\dag}_{n,tr}|0\ket$ is
unphysical. This statement is almost obvious when one recalls what
each creation operator does.

We should however note that a true physical state be summed over all
the transverse sites since we have discrete translational symmetry
in the transverse  direction. That is, for example, the states
$d^{I\dag}_{11,rs}a^{I\dag}_{11,sr}|0\ket$ and
$d^{I\dag}_{12,rs}a^{I\dag}_{12,sr}|0\ket$ are the same up to a
phase factor given by $\exp(iP^{2}a)$. We set the phase factor to
one since we take physical states to have $P^{1}=P^2=0$.   The
physical state $|phys \ 1\ket$ is in fact
$\sum_{i,j=1}^{N_{sites}}d^{I\dag}_{ij,rs}a^{I\dag}_{ij,sr}|0\ket$
with the appropriate normalization constant. From a computational
point of view this leads to a great simplification in the large
$N_c$ limit. Because as shown in appendix \ref{reduction} it turns
out that in the large $N_c$ limit we can drop the site index $n$
from the expression of the supercharges and thus can practically set
$N_{sites}=1$ for our calculation. This is in some sense the
manifestation of the Eguchi-Kawai reduction \cite{egk83}.
Eguchi-Kawai reduction tells us in the usual lattice theory that the
large $N_c$ limit allows us to work with only one site in each of
the space-time directions in Euclidean space. However, the way we
justify this reduction in our transverse lattice formulation is
quite different from the way Eguchi and Kawai do in the usual
lattice formulation. Therefore, we believe that we do not have to
introduce quenching \cite{Bhanot:1982sh} or twisted
\cite{Gonzalez-Arroyo:1982hz} lattices to overcome the problem that
the naive Eguchi-Kawai reduction comes across at weak couplings
\cite{Bhanot:1982sh}. We refer the reader to appendix
\ref{reduction} for more detailed support for this claim.

Periodic conditions on the fields allow for physical states of the
form $|phys \ 2\ket \equiv \sum_n
a^{I\dag}_{n+(N_{sites}-1)i_I,ru}\cdots a^{I\dag}_{n+i_I,ts}
a^{I\dag}_{n,sr}|0\ket$. The color for this state is carried around
the transverse lattice, as shown in Fig.~\ref{da}b.  We will refer
to these states as cyclic states. The states where the color flux
does not go all the way around the transverse lattice we will refer
to as non-cyclic states. We characterize  states by what we call the
winding number defined by $W_I=n_I/N_{sites}$, where $n_I\equiv
\sum_n(a_n^{I\dag}a_n^I-d^{I\dag}_nd_n^I)$. For $N_{sites} = 1$, the
winding number $W_I$ simply gives us the excess number of
$a^{I\dag}$ over $d^{I\dag}$ in a state. We use the winding number
to classify states since the winding number is a good quantum number
commuting with $P^-_{SDLCQ}$ as we will see in the next section. In
the language of the winding number the non-cyclic states  are those
states with $W_I=0$ and cyclic states have non-zero $W_I$.

\section{SDLCQ of the transverse lattice model}
The transverse lattice formulation of ${\cal N}=1$ SYM theory in 3+1
dimension presented in the previous section has several undesirable
features. First and foremost the naive Lagrangian suffers from the
fermion species doubling problem \cite{Harada:2004cc}. Second, the
supersymmetric structure of the theory is completely hidden. Lastly,
the resulting Hamiltonian is $6^{th}$ order in the dynamical fields.
From the numerical point of view a $6^{th}$ order interaction makes
the theory considerably more difficult to solve. In
Ref.~\cite{Harada:2003bs} we found that the (2+1)--dimensional
supersymmetric Hamiltonian is only $4^{th}$ order making this
discrete formulation of the theory very different. Unfortunately, it
seems this is not the case for 3+1 dimensional model. Instead we
seem to have supersymmetric Hamiltonian of $8^{th}$ order in fields.
However, since this SDLCQ Hamiltonian is free from the doubling
problem \cite{Harada:2004cc} and since the supercharge $Q^-_{\a}$,
where $\a=1,2$, is of $5^{th}$ order and it is this $Q^-_{\a}$ that
we make use of for our calculations, we think that this SDLCQ
formulation is still more advantageous than the naive DLCQ
formulation. There can, of course, be many discrete formulations
that correspond to the same continuum theory and it is therefore
desirable to search for a better one.

In the spirit of SDLCQ we will attempt a discrete formulation based
on the underlying super-algebra of this theory,
\begin{equation}
 \{Q^{\pm}_{\a},Q^\pm_{\beta}\}=2\sqrt 2 P^\pm\delta_{\a\beta},
 \quad    \{Q^+_{\a},Q^-_{\beta}\}=0, \label{susyalgebra}
\end{equation}
where $\a,\beta=1,2$ and the supercharge $Q$ is given by
\[
   Q\equiv \sum_{n}\int dx^- j^+_n\equiv\left(
   \begin{array}{c}Q^+_1\\Q^+_2 \\ Q^-_1\\Q^-_2 \end{array}\right)
\]
with $j^{\mu}_n$ being the supercurrent at the site $n=(i,j)$, which
is a Majorana spinor. For the derivation of the super-algebra in
Majorana representation Eq.~\eqref{susyalgebra}, see the appendix
\ref{majorana}. We've set $P^I=0$ with $I=1,2$ since we're
considering the physical states only with $P^I=0$. Note that this
choice of $P^I$ has made Eq.~\eqref{susyalgebra} coincide with the
${\cal N}$=2 super-algebra in 1+1 dimensions also known as ${\cal
N}$=(2,2) super-algebra although we are considering ${\cal N}$=1 SYM
in 3+1 dimensions.

In this effort however there are some fundamental limits to how far
one can go. As we discussed in the previous section the physical
states of this theory must conserve the color charge at every point
on the transverse lattice. Experience with other supersymmetric
theories indicates that each term in $Q^+_{\a}$ has to be either the
product of {\it one} $M_n$ and {\it one} $\psi_n$ or of {\it one}
$M_n^{\dag}$ and {\it one} $\psi_n$ therefore $Q^+_{\a}$ is {\it
unphysical}, by which we mean that $Q^+_{\a}$ transforms a physical
state into an unphysical one, so that $\bra phys|Q^+_{\a}|phys
\ket=0$. While this is not a theorem, it seems very difficult to
have any other structure since in light cone quantization $P^+$ is a
kinematic operator and therefore independent of the coupling. There
appears to be  {\it no} way to make a physical $P^+$ from
$Q^+_{\a}$. We will use $P^+$ as given in Eq. \eqref{pplus} in what
follows. Similarly, we are not able to generally construct physical
$P^{I}$ from $Q^+_{\a}$ and $Q^-_{\beta}$. In fact $P^{I}$ is
unphysical in our formalism, leading to $\bra phys|P^I|phys \ket=0$.
Formally we will work in the frame where total $P^I$ is zero, so it
would appear consistent with this result. We should note, however,
that this is not totally satisfying because $P^I=0$ was a choice and
a non-zero value is equally valid and not consistent with the matrix
element.

Despite these difficulties we find a physical $Q^-_{\a}$ which gives
us $P^-_{SDLCQ}\stackrel{a\to 0}{\longrightarrow} P^-_{cont}$. The
expression for $Q^-_{\a}$ is

\begin{eqnarray*}
  Q_{\a}^- &=&i2^{3/4}a^2\sum_{n}\int dx^- \tr\Biggl\{ \Bigg[
  \frac{-i}{2ga^2}\left(M_n^I\stackrel{\leftrightarrow}{\d_-}M_n^{I\dag}
  +M_{n-i_I}^{I\dag}\stackrel{\leftrightarrow}{\d_-}M_{n-i_I}^I\right)
  -2g\psi_{Rn}^T\psi_{Rn}\Bigg]\\
  && \times \frac 1{\d_-}(\sigma_2\psi_{Rn})_{\a}
  +\frac{-i}{2ga^2}(M_n^IM_{n+i_I}^JM_{n+i_J}^{I\dag}M_n^{J\dag}-1)
    (\beta_I\beta_J\sigma_2\psi_{Rn})_{\a}\Biggr\}\\
  &&\stackrel{a\to 0}{\longrightarrow}i2^{-1/4}\int d^3x
  \left\{-2gJ^+\frac 1{\d_-}(\sigma_2\psi_R)_{\a}
  +F_{IJ}(\beta_I\beta_J\sigma_2\psi_R)_{\a}\right\},
\end{eqnarray*}
where $\beta_1\equiv \sigma_1$, $\beta_2=\sigma_3$, $gJ^+\equiv
ig[A_{I},\d_-A_{I}]+\d_{I}\d_-A_{I}+2g\psi_R\psi_R$, and the last
line is the continuum form for $Q^-_{\a}$ in 3+1 dimensions.

It is tedious but straightforward to check that $\{Q^-_1,Q^-_1\}\ne
\{Q^-_2,Q^-_2\}$, while both $\{Q^-_1,Q^-_1\}$ and $\{Q^-_2,Q^-_2\}$
give the same correct $P^-$ in the limit of $a \to 0$. In addition,
one can show that $\{Q^-_1,Q^-_2\}\ne 0$ in the discrete form but
becomes zero as $a\to 0$. This means that we preserve {\it only one}
supersymmetry algebra, say $\{Q^-_1,Q^-_1\}=P^-$, in our discrete
formalism. We cannot use both $Q^-_1$ and $Q^-_2$ at the same time
to construct physical states since they do not commute with each
other. However, both $Q^-_1$ and $Q^-_2$ separately give us the same
mass spectrum when we perform SDLCQ calculations. Thus, it is
suffice to consider only one of the two and we take $Q^-_1$ for our
calculations in the following sessions.

 Notice that $Q^-_{\a}$ above is fifth order and, thus,
$P^-_{SDLCQ}$ obtained from it is eighth order in fields as we
mentioned at the beginning of this section. In fact we find

\begin{eqnarray*}
&&P^-_{SDLCQ}\equiv \frac{\{Q_1^-,Q_1^-\}}{2\sqrt 2}\\
&&=a^2\sum_n\int dx \tr \Bigg\{
  -\frac{g^2}2 \frac{i}{2g^2a^2}\left[\left(M_n^I\stackrel{\leftrightarrow}{\d}
  M_n^{I\dag}+M_{n-i_I}^{I\dag}\stackrel{\leftrightarrow}{\d}
  M_{n-i_I}^I\right)
  +2u^{\a}_{n}u^{\a}_{n}\right]  \\
&& \qquad \times
  \frac 1{\d^2}\frac{i}{2g^2a^2} \left[
  \left(M_n^I\stackrel{\leftrightarrow}{\d}M_n^{I\dag}
  +M_{n-i_I}^{I\dag}\stackrel{\leftrightarrow}{\d}M_{n-i_I}^I\right)
  +2u^{\a}_{n}u^{\a}_{n}\right]   \\
&&-\frac {i}{2a^2}(u^2_{n+i_I}M_{n}^{I\dag}-M_{n}^{I\dag}u^2_n)
    \d^{-1}(M_n^Iu^2_{n+i_I}-u^2_nM_n^I) +\frac {i}{2a^2}\Big\{\\
&& (u^2_{n+i_1+i_2} M_{n+i_2}^{1\dag}-M_{n+i_2}^{1\dag}
  u^2_{n+i_2})\d^{-1}(M_{n}^{2\dag}
  u^1_{n}M_{n}^1M_{n+i_1}^2-u^1_{n+i_2}
  M_{n+i_2}^2M_{n+2i_2}^1M_{n+i_1+i_2}^{2\dag}) \\
&& +(u^2_{n+i_2} M_{n}^{2\dag}-M_{n}^{2\dag}u^2_{n})
  \d^{-1}(u^1_{n}M_{n}^1M_{n+i_1}^2M_{n+i_2}^{1\dag}
  -M_{n-i_1}^{1\dag} u^1_{n-i_1}M_{n-i_1}^2
  M_{n-i_1+i_2}^1) \\
&& +(u^2_{n+i_1}M_{n+i_1}^2 -M_{n+i_1}^{2}
  u^2_{n+i_1+i_2})\d^{-1}
  (M_{n+i_2}^{1\dag}M_{n}^{2\dag}u^1_{n}M_{n}^1
  -M_{n+i_1+i_2}^1M_{n+2i_1}^{2\dag}M_{n+i_1}^{1\dag}
  u^1_{n+i_1}) \\
&& +(u^2_{n} M_{n}^1-M_{n}^{1}u^2_{n+i_1})
  \d^{-1}(M_{n+i_1}^2M_{n+i_2}^{1\dag}M_{n}^{2\dag}
  u^1_{n}-M_{n+i_1-i_2}^{2\dag}M_{n-i_2}^{1\dag}
  u^1_{n-i_2}M_{n-i_2}^2) \Big\} \\
&& -\frac{i}{4a^2}\Bigl\{
  (M_{n+i_1}^2M_{n+i_2}^{1\dag}M_{n}^{2\dag}u^1_{n}
  -M_{n+i_1-i_2}^{2\dag}M_{n-i_2}^{1\dag}
  u^1_{n-i_2}M_{n-i_2}^2) \\
&& \times \d^{-1}(u^1_{n}M_{n}^2M_{n+i_2}^1M_{n+i_1}^{2\dag}
  -M_{n-i_2}^{2\dag}u^1_{n-i_2}M_{n-i_2}^1
  M_{n+i_1-i_2}^2) \\
&& +(M_{n}^{2\dag}u^1_{n}M_{n}^1M_{n+i_1}^2
  -u^1_{n+i_2}M_{n+i_2}^2
  M_{n+2i_2}^1M_{n+i_1+i_2}^{2\dag}) \\
&& \times
  \d^{-1}(M_{n+i_1}^{2\dag}M_{n}^{1\dag} u^1_{n}M_{n}^2
  -M_{n+i_1+i_2}^2M_{n+2i_2}^{1\dag}
  M_{n+i_2}^{2\dag}u^1_{n+i_2})\\
&& +(M_{n+i_2}^{1\dag}M_{n}^{2\dag}u^1_{n}M_{n}^1
  -M_{n+i_1+i_2}^1M_{n+2i_1}^{2\dag}M_{n+i_1}^{1\dag}
  u^1_{n+i_1})  \\
&& \times
  \d^{-1}(M_{n}^{1\dag}u^1_{n}M_{n}^2M_{n+i_2}^1
  -u^1_{n+i_1}M_{n+i_1}^1M_{n+2i_1}^2
  M_{n+i_1+i_2}^{1\dag}) \\
&& +(u^1_{n}M_{n}^1M_{n+i_1}^2M_{n+i_2}^{1\dag}
  -M_{n-i_1}^{1\dag}u^1_{n-i_1}M_{n-i_1}^2
  M_{n-i_1+i_2}^1  \\
&& \times
  \d^{-1}(M_{n+i_2}^1M_{n+i_1}^{2\dag}M_{n}^{1\dag}u^1_{n}
  -M_{n-i_1+i_2}^{1\dag}M_{n-i_1}^{2\dag}
  u^1_{n-i_1}M_{n-i_1}^1)  \Big\} \\
&& +\frac{-1}{8a^2}
  (M_n^1M_{n+i_1}^2M_{n+i_2}^{1\dag}M_n^{2\dag}
   -M_n^2M_{n+i_2}^1M_{n+i_1}^{2\dag}M_n^{1\dag})^2\Bigg\}.
\end{eqnarray*}

One can show that by setting $g=0$ and $M,M^{\dag}=1$ this
$P^-_{SDLCQ}$ gives rise to a dispersion relation
\[ k^-=\frac 1{2k^+}\left[ \left(\frac{\sin\frac{k^1a}{2}}{a/2}
\right)^2+\left(\frac{\sin\frac{k^2a}{2}}{a/2} \right)^2\right],
\]
which is free from the fermion species doubling problem
\cite{Harada:2004cc}. Furthermore, one can check that this $Q^-$
commutes with $P^+$ obtained from ${\cal L}$; $[Q^-,P^+]=0$. Thus,
it follows that,
\begin{equation}
  \bra phys|[Q^-,M^2]|phys\ket =\bra phys|[Q^-,2P^+P^-_{SDLCQ}]|phys\ket =0
  \label{supersymmetry}
\end{equation}
in our SDLCQ formalism, where $M^2 \equiv
2P^+P^-_{SDLCQ}-(P^1)^2-(P^2)^2$. The fact that the Hamiltonian is
the square of a supercharge will guarantee the usual supersymmetric
degeneracy of the massive spectrum, and our numerical solutions will
substantiate this. Unfortunately one needs a $Q^+$ to guarantee the
degeneracy of the massless bound states.

Recalling that we set $N_{sites}=1$ in both transverse directions
and that we are in the large-$N_c$ limit. we can write $Q^-_1$ as

\[ Q^-_1={\cal Q}^-_{11}+{\cal Q}^-_{12}+{\cal Q}^-_{13},\]
where
\begin{eqnarray}
&&{\cal Q}^-_{11}=-\frac{ i2^{-1/4}a^2g}{\sqrt{\pi}} \int_0^{\infty}
    dk_1dk_2dk_3\delta(k_1+k_2-k_3)\nonumber\\
&&\times\Big[
 \frac{k_2-k_1}{k_3\sqrt{k_1k_2}}
   (-b^{2\dag}d^{I}a^I+d^{I\dag}a^{I\dag}b^{2}
   -b^{2\dag}a^Id^{I} +a^{I\dag}d^{I\dag}b^{2}) \nonumber\\
&&+\frac{k_2+k_3}{k_1\sqrt{k_2k_3}}
   (-d^{I\dag}b^{2}d^I+b^{2\dag}d^{I\dag}d^{I}
   -a^{I\dag}b^2a^{I} +b^{2\dag}a^{I\dag}a^{I}) \nonumber\\
&&+\frac{k_3+k_1}{k_2\sqrt{k_3k_1}}
   ( a^{I\dag}a^{I}b^2-a^{I\dag}b^{2\dag}a^{I}
   +d^{I\dag}d^Ib^{2}-d^{I\dag}b^{2\dag}d^{I})\nonumber\\
&&   +\left(\frac 1{k_1}+\frac 1{k_2}-\frac 1{k_3}\right)
    (b^{2\dag}b^{2\dag}b^{2}+b^{2\dag}b^{2}b^{2})\Big]\label{q1},
\end{eqnarray}
\begin{eqnarray}
&&{\cal Q}^-_{12}=-\frac{ i2^{-1/4}a^2g}{\sqrt{\pi}} \int_0^{\infty}
    dk_1dk_2dk_3\delta(k_1+k_2-k_3)\nonumber\\
&& \times\Big[ \frac{-1}{k_3}(b^{2\dag}b^1b^1+b^{1\dag}b^{1\dag}b^2)
  +\frac 1{k_1}(b^{1\dag}b^2b^1+b^{2\dag}b^{1\dag}b^1)
  +\frac 1{k_2}(b^{1\dag}b^{2\dag}b^1+b^{1\dag}b^1b^2)\Big]\label{q2},
\end{eqnarray}
\begin{eqnarray}
&&{\cal Q}^-_{13}=-\frac{i2^{-1/4}a^2g}{\sqrt{\pi}} \frac
{a^2g^2}{4\pi}
 \int_0^{\infty}dk_1dk_2dk_3dk_4dk_5  \Big\{ \delta(k_1+k_2+k_3+k_4-k_5)\Big[
 \nonumber\\
&& \frac 1{\sqrt{k_1k_2k_3k_4}}
  (b^{1\dag}d^1d^2a^1a^2+d^{2\dag}d^{1\dag}a^{2\dag}a^{1\dag}b^1
  -b^{1\dag}d^2d^1a^2a^1-d^{1\dag}d^{2\dag}a^{1\dag}a^{2\dag}b^1)
  \nonumber \\
&& +\frac 1{\sqrt{k_2k_3k_4k_5}}
  (d^{2\dag}b^{1}d^1d^2a^1+b^{1\dag}d^{2\dag}d^{1\dag}a^{2\dag}d^{1}
  -d^{1\dag}b^{1}d^2d^1a^2-b^{1\dag}d^{1\dag}d^{2\dag}a^{1\dag}d^{2})
   \nonumber\\
&& +\frac 1{\sqrt{k_3k_4k_5k_1}}
  (d^{1\dag}a^2b^{1}d^1d^2+a^{1\dag}b^{1\dag}d^{2\dag}d^{1\dag}d^{2}
  -d^{2\dag}a^1b^{1}d^2d^1-a^{2\dag}b^{1\dag}d^{1\dag}d^{2\dag}d^{1})
  \nonumber\\
&& +\frac 1{\sqrt{k_4k_5k_1k_2}}
  (a^{2\dag}a^1a^2b^{1}d^1+a^{2\dag}a^{1\dag}b^{1\dag}d^{2\dag}a^{1}
  -a^{1\dag}a^2a^1b^{1}d^2-a^{1\dag}a^{2\dag}b^{1\dag}d^{1\dag}a^{2} )
  \nonumber\\
&& +\frac 1{\sqrt{k_5k_1k_2k_3}}
  (a^{1\dag}d^2a^1a^2b^{1}+d^{1\dag}a^{2\dag}a^{1\dag}b^{1\dag}a^{2}
  -a^{2\dag}d^1a^2a^1b^{1}-d^{2\dag}a^{1\dag}a^{2\dag}b^{1\dag}a^{1}) \Big]
  \nonumber \\
&& +\delta(k_1+k_2+k_3-k_4-k_5)\nonumber\\
&& \times\Big[
  \frac 1{\sqrt{k_1k_2k_3k_4}}
  (d^{1\dag}a^{2\dag}a^{1\dag}a^{2}b^{1}+a^{1\dag}b^{1\dag}d^{2}a^{1}a^{2}
  -d^{2\dag}a^{1\dag}a^{2\dag}a^{1}b^{1}-a^{2\dag}b^{1\dag}d^{1}a^{2}a^{1})
  \nonumber\\
&& +\frac 1{\sqrt{k_2k_3k_4k_5}}
  (b^{1\dag}d^{2\dag}d^{1\dag}d^{1}d^2+d^{2\dag}d^{1\dag}b^{1}d^{1}d^{2}
  -b^{1\dag}d^{1\dag}d^{2\dag}d^{2}d^1-d^{1\dag}d^{2\dag}b^{1}d^{2}d^{1})
  \nonumber\\
&& +\frac 1{\sqrt{k_3k_4k_5k_1}}
  (a^{1\dag}b^{1\dag}d^{2\dag}d^2a^{1}+d^{1\dag}a^{2\dag}a^{2}b^{1}d^{1}
  -a^{2\dag}b^{1\dag}d^{1\dag}d^1a^{2}-d^{2\dag}a^{1\dag}a^{1}b^{1}d^{2})
  \nonumber\\
&& +\frac 1{\sqrt{k_4k_5k_1k_2}}
  (a^{2\dag}a^{1\dag}b^{1\dag}a^{1}a^{2}+a^{2\dag}a^{1\dag}a^{1}a^{2}b^{1}
  -a^{1\dag}a^{2\dag}b^{1\dag}a^{2}a^{1}-a^{1\dag}a^{2\dag}a^{2}a^{1}b^{1})
  \nonumber\\
&&  +\frac 1{\sqrt{k_5k_1k_2k_3}}
  (d^{2\dag}d^{1\dag}a^{2\dag}b^{1}d^{1}+b^{1\dag}d^{2\dag}d^{1}d^{2}a^{1}
  -d^{1\dag}d^{2\dag}a^{1\dag}b^{1}d^{2}-b^{1\dag}d^{1\dag}d^{2}d^{1}a^{2})
  \Big]\Big\}\label{q3},
\end{eqnarray}
with $k^+ \equiv k$, $a_1\equiv a(k_1)$, $a^{\dag}aa\equiv
\tr(a^{\dag}_3a_1a_2)$, $a^{\dag}a^{\dag}a\equiv
\tr(a^{\dag}_1a^{\dag}_2a_3)$, $a^{\dag}aaaa\equiv
\tr(a^{\dag}_5a_1a_2a_3a_4)$,
$a^{\dag}a^{\dag}a^{\dag}a^{\dag}a\equiv
\tr(a^{\dag}_1a^{\dag}_2a^{\dag}_3a^{\dag}_4a_5)$,
$a^{\dag}a^{\dag}a^{\dag}aa\equiv
\tr(a^{\dag}_1a^{\dag}_2a^{\dag}_3a_4a_5)$, and
$a^{\dag}a^{\dag}aaa\equiv \tr(a^{\dag}_4a^{\dag}_5a_1a_2a_3)$.
${\cal Q}_{11}$ is the part of $Q^-_1$ which looks exactly like
$Q^-$ in 2+1 dimensional model with the difference being that here
we have two types for each of the bosonic fields $a$ and $d$. ${\cal
Q}_{12}$ is a new piece in 3+1 dimensions and mixes two different
types of fermionic fields. ${\cal Q}_{13}$ is also new and composed
of fields of fifth order. Note that for small couplings, ${\cal
Q}_{11}$ and ${\cal Q}_{12}$ dominate over ${\cal Q}_{13}$, while
${\cal Q}_{13}$ dominates in the strong coupling regime. Notice that
from this explicit expression for $Q^-_1$ it is clear that the
winding number introduced in the last section evidently commutes
with $Q^-_1$ and, thus, with $P^-_{SDLCQ}$. Therefore, cyclic states
do not mix with non-cyclic states.

It is always important to look for symmetries of $Q^-$ since the
symmetries, if any, will reduce the amount of the computational
efforts considerably. To do this, let us consider three cases
separately: (i) the intermediate coupling where we have all the
three pieces together for $Q^-_1$; (ii) the weak coupling limit
where we can ignore ${\cal Q}_{13}$; (iii) the strong coupling limit
where we consider ${\cal Q}_{13}$ only. For the first case (i) we
find two $Z_2$ symmetries,
\begin{itemize}
\item $a^1_{ij}\leftrightarrow -a^2_{ij}, \
d^1_{ij}\leftrightarrow -d^2_{ij}, \ b^1_{ij}\leftrightarrow
-b^1_{ij}, \ b^2 \ {\rm unchanged}$,
\item $a^I_{ij}\leftrightarrow -d^I_{ji}, \
b^{\a}_{ij}\leftrightarrow -b^{\a}_{ji}$.
\end{itemize}

The first symmetry implies that states with the winding numbers, say
$(W_1,W_2)$, are equivalent to those with $(W_2,W_1)$ up to the
minus sign. On the other hand the second symmetry implies that
states with $(W_1,W_2)$ are equivalent to those with $(-W_1,-W_2)$
up to the minus sign.

In the case of the weak coupling limit (ii), we find two more
independent $Z_2$ symmetries;
\begin{itemize}
\item $a^I_{ij}\leftrightarrow -a^I_{ji}, \
d^I_{ij}\leftrightarrow -d^I_{ji}, \ b^{\a}_{ij}\leftrightarrow
-b^{\a}_{ji}$.
\item $a^1_{ij}\leftrightarrow -d^1_{ji}, \
a^2_{ij}\leftrightarrow -a^2_{ji}, \ d^2_{ij}\leftrightarrow
-d^2_{ji}, \ b^{\a}_{ij}\leftrightarrow -b^{\a}_{ji}$.
\end{itemize}

The second of these implies, with the help of the second $Z_2$
symmetry we found in the case of (i), the equivalence of states
under $(W_1,W_2)\leftrightarrow (-W_1,W_2)\leftrightarrow
(W_1,-W_2)$.

In the strong coupling limit (iii), we do not have any other $Z_2$
symmetries besides the two we found in the case of (i). However, it
is easy to see that ${\cal Q}_{13}$ commutes with $b^{2\dag}b^2$,
thus the number of $b^{2\dag}$'s is a good quantum number as well as
the two winding numbers.

It is interesting to see what we can find for each of the three
different cases (i), (ii) and (iii). However, in this our first
attempt to formulate ${\cal N}=(2,2)$ SYM in 3+1 dimensions with
SDLCQ on a two dimensional transverse lattice, we constrain
ourselves to consider only the most generic case (i) where we have
all the three pieces together for $Q^-$.

 Now we are in a position to solve the eigenvalue problem $
2P^+P^-_{SDLCQ}|phys\ket=m^2|phys\ket$.   We impose the periodicity
condition on $M^I_n$, $M^{I\dag}_n$ and $u^{\a}_n$ in the $x^-$
direction giving a discrete spectrum for $k^+$, and ignore the
zero-mode:
\[
k^+=\frac {\pi}L n \quad (n=1,2,\ldots.), \qquad \int_0^{\infty}
dk^+
   \rightarrow \frac {\pi}L \sum_{n=1}^{\infty}.
\]
We impose a cut-off on the total longitudinal momentum $P^+$ i.e.
$P^+=\pi K/L$, where $K$ is an integer also known as the `harmonic
resolution', which indicates the coarseness of our numerical
results.  For a fixed $P^+$ i.e. a fixed $K$, the number of partons
in a state is limited up to the maximum, that is $K$, so that the
total number of Fock states is {\it finite}, and, therefore, we have
reduced the infinite dimensional eigenvalue problem to a finite
dimensional one.

For this initial study of the transverse lattice we consider
resolution up to $K=8$ for non-cyclic ($W_1=W_2=0$) states and up to
$K=W_I+6$ and $K=W_I+5$ for states with $|W_I|=1$ and
$|W_I|=2,3,4,5$, respectively. We were able to handle these
calculations with our SDLCQ Mathematica code and $C++$ code.

\section{Coupling dependence of the mass spectrum }
In this section we will discuss the mass spectrum as a function of
$g'\equiv g\sqrt{N_c}$ for $K=4,5,6$.

\begin{figure}[t!]
\begin{tabular}{cc}
\epsfig{figure=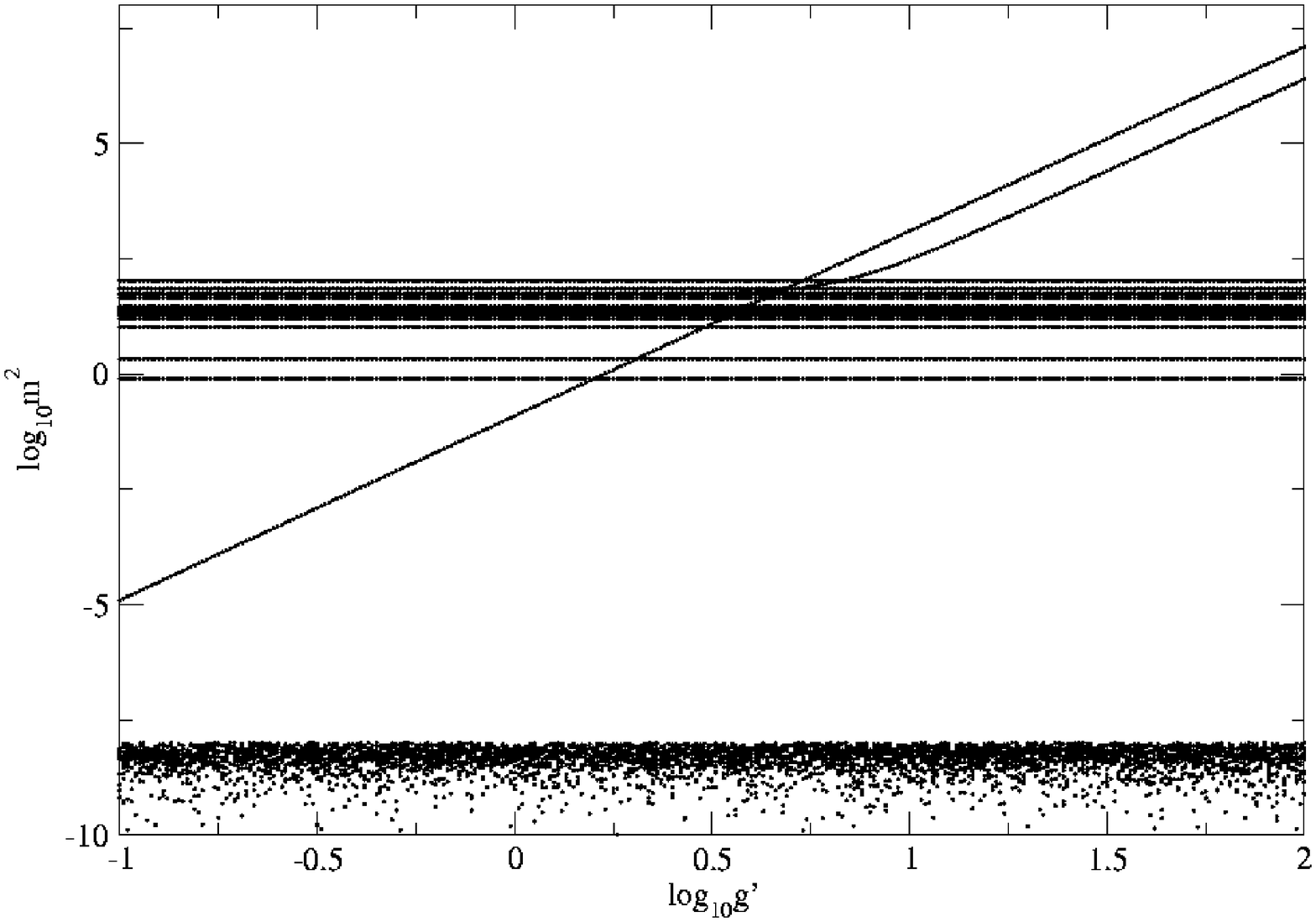,width=7.3cm,angle=0}&
\epsfig{figure=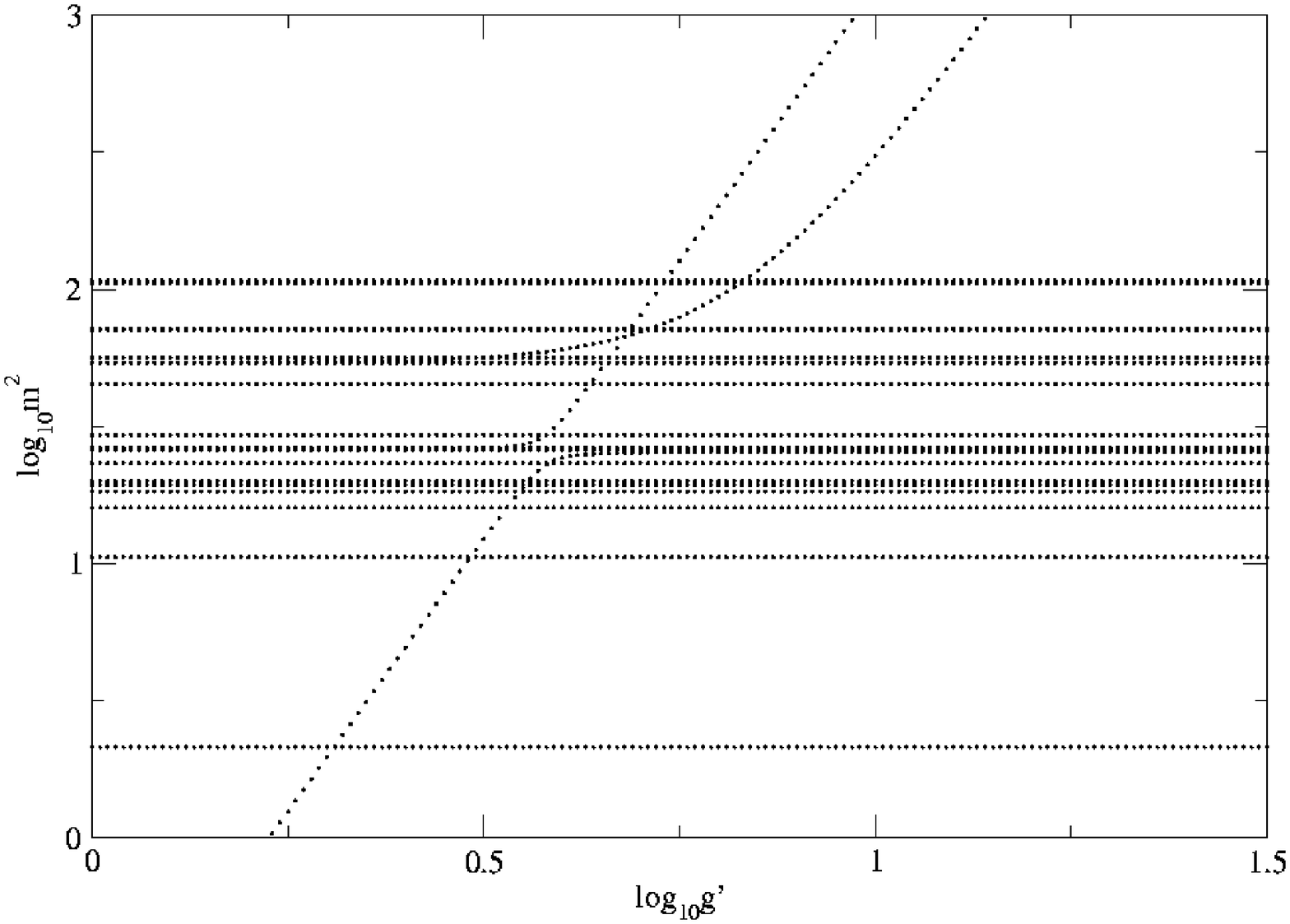,width=7.3cm,angle=0}\\
\vspace{1cm}
(a)&(b)\\
\epsfig{figure=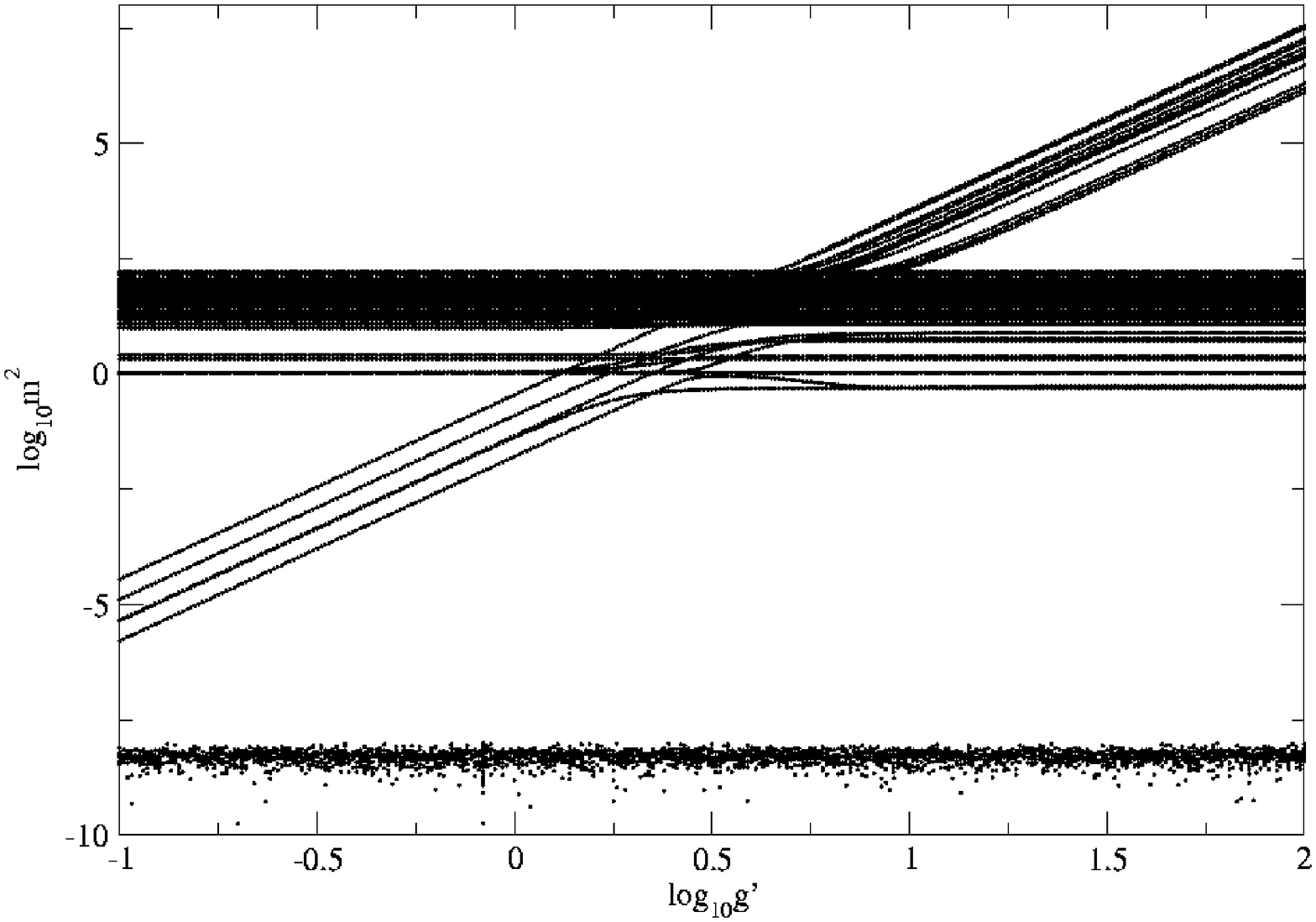,width=7.3cm,angle=0}&
\epsfig{figure=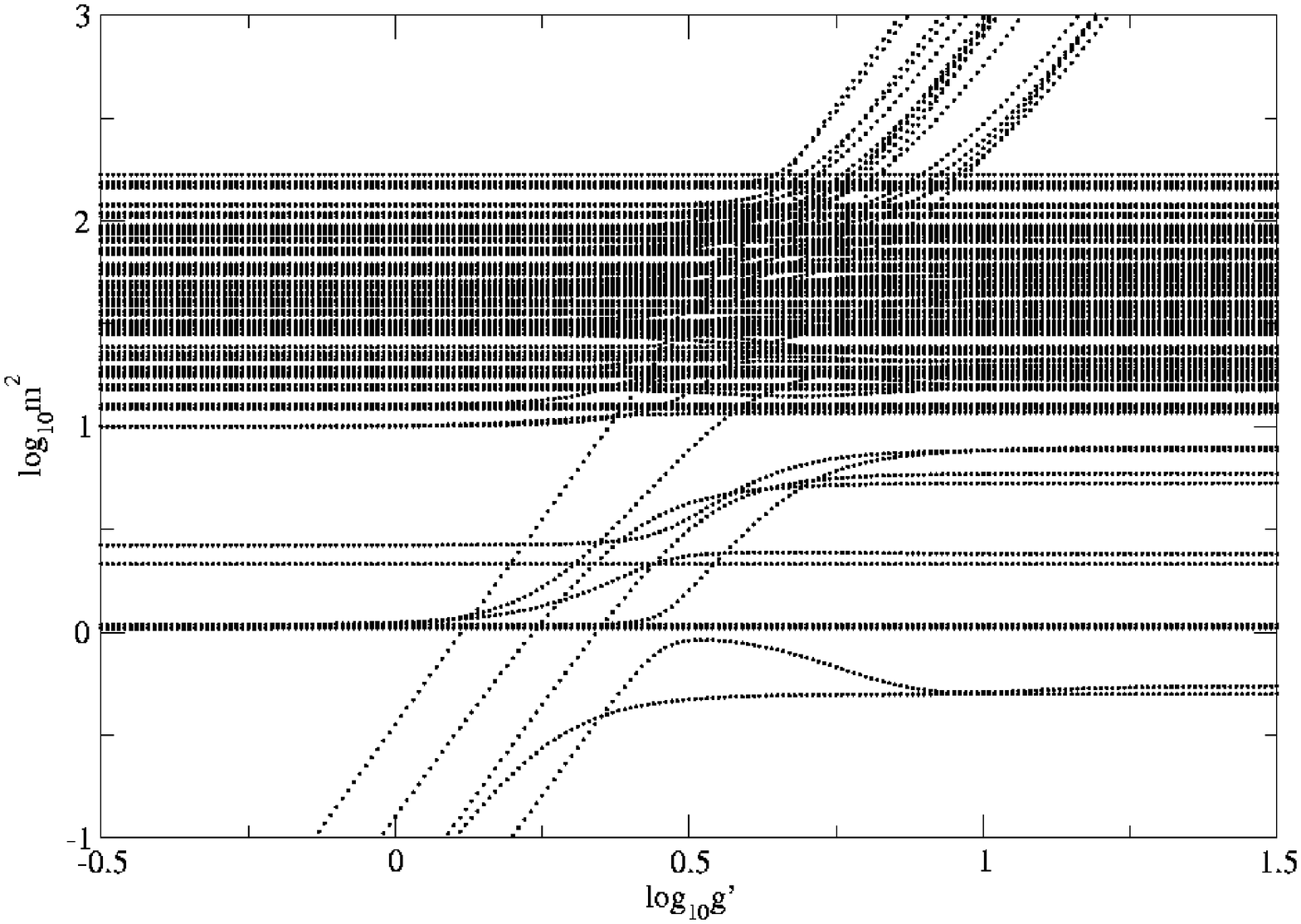,width=7.3cm,angle=0}\\
\vspace{1cm}
(c)&(d)\\
\epsfig{figure=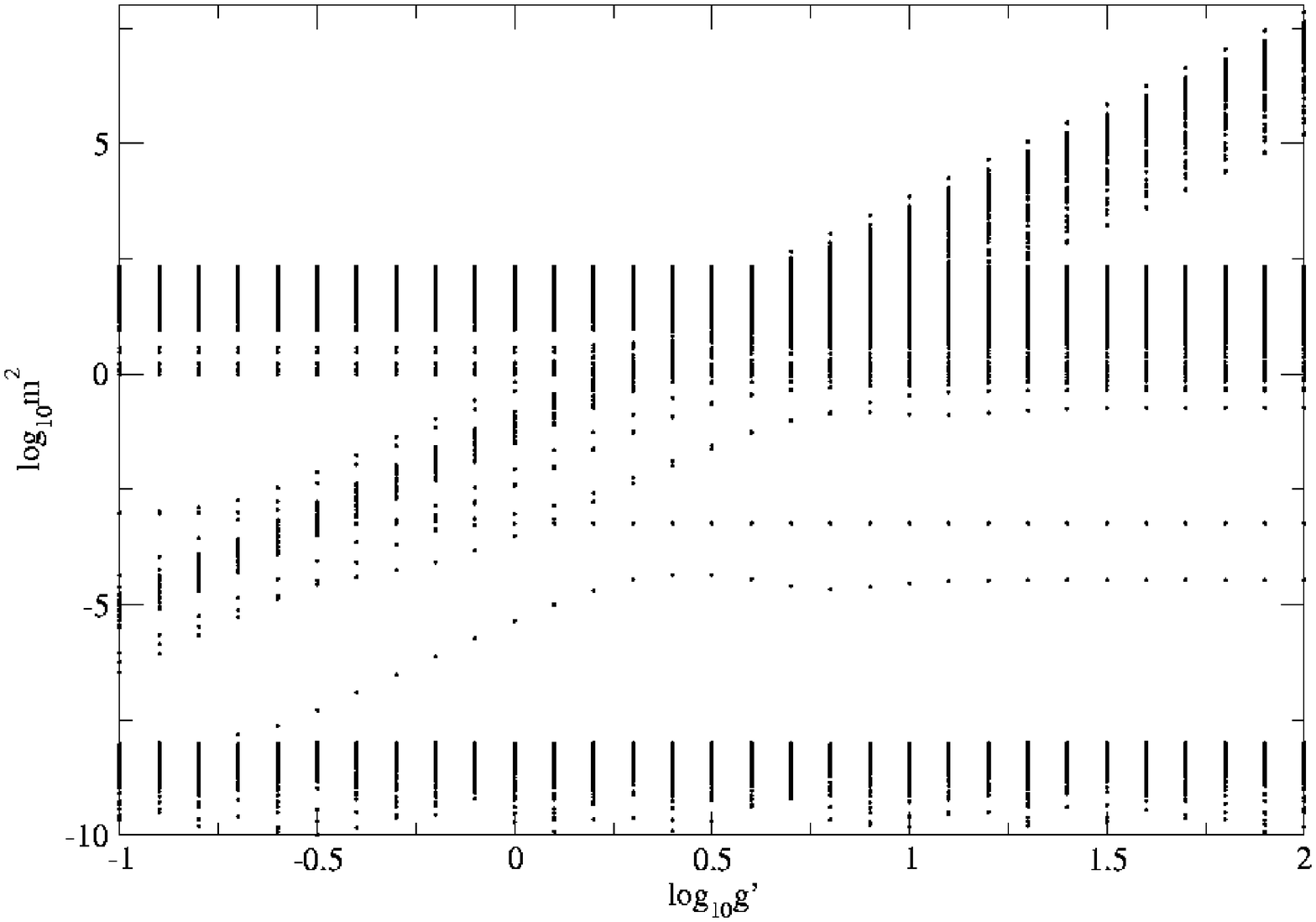,width=7.3cm,angle=0}&
\epsfig{figure=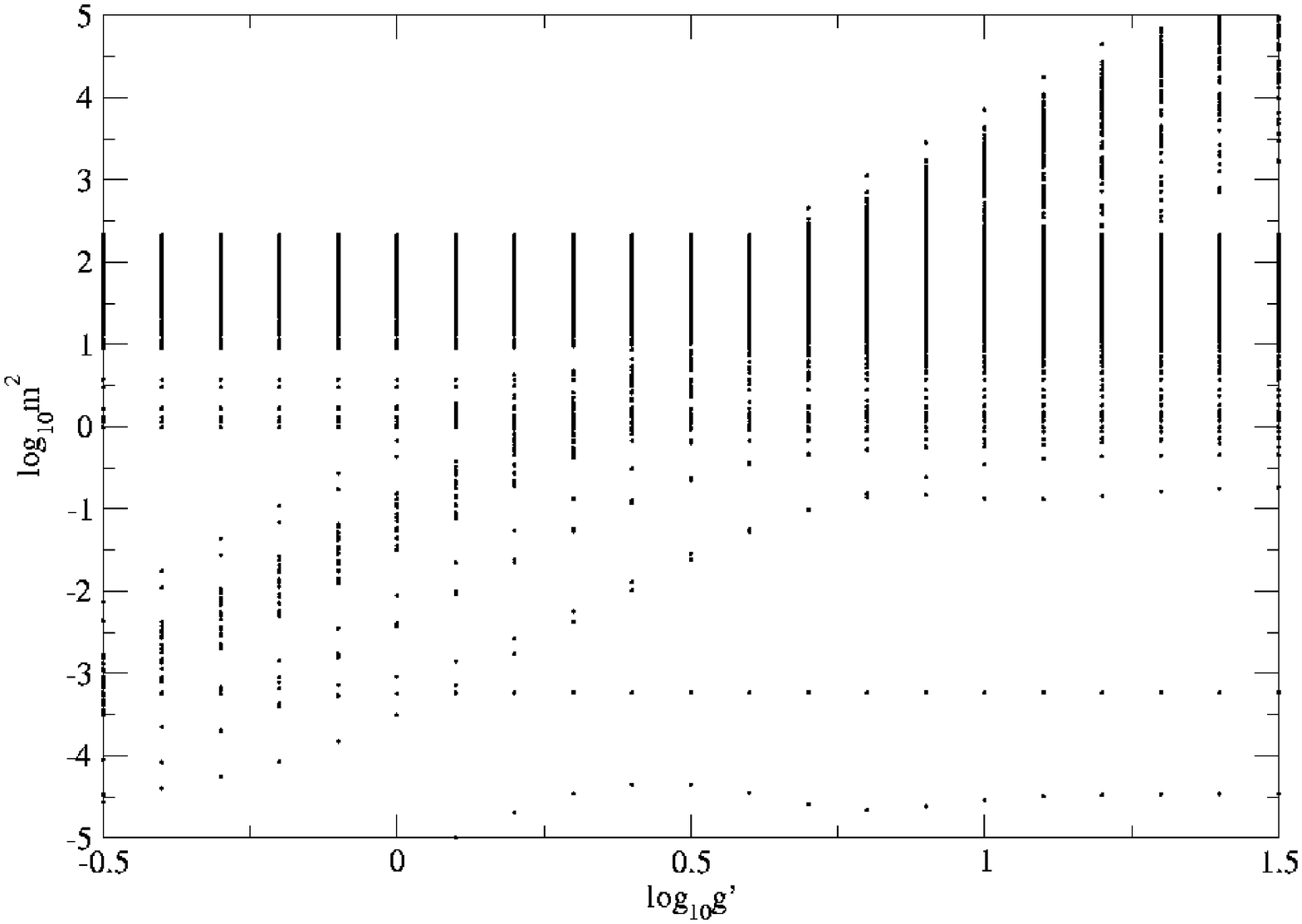,width=7.3cm,angle=0}\\
(e)&(f)
\end{tabular}

\caption{Log-log plots of the mass spectrum $m^2$ in units of
$\frac{g'^{2}}{\pi a^2}$ versus $g'\equiv g\sqrt {N_c}$ with
$K=4,5,6$ for (a),(c),(e), respectively. (b), (d), and (f) are the
same as (a), (c) and (e), respectively but on a different scale so
that one can see the crossings in more detail. $10^{-8}$ or less is
the numerical zero in our code.} \label{spectrum}
\end{figure}

It is instructive to see the dependence of $m^2$ on the coupling
since we have terms in $Q^-$ that go like $g'$ and $g'^3$. In
Fig.~\ref{spectrum} we show the entire mass spectrum of non-cyclic
states in units of $g'^2/\pi a^2$ for $K=4,5,6$ as a function of
$g'$ in a log-log plot. In order to see the crossings in more detail
we show Fig.~\ref{spectrum}(b), (d), and (f) on a different scale
from (a), (c) and (e), respectively. We've set $10^{-8}$ or less to
the numerical zero in our code.

As one can see from Fig.~\ref{spectrum}, there is a rich structure
in the mass spectrum as a function of $g'$, and the origin of this
structure for the case where $K=4$ in Fig.~\ref{spectrum}(a) and (b)
is rather easy to understand. We find four types of states; (i)
those states which are killed by ${\cal Q}_{13}$ and whose $m^2$ in
units of $\frac{g'^2}{\pi a^2}$ are independent of $g'$; (ii) those
states which vanish upon the action of ${\cal Q}_{11}+{\cal Q}_{12}$
and thus whose $m^2$ in units of $\frac{g'^2}{\pi a^2}$ go like
$g'^4$; (iii) those states which survive upon the action of ${\cal
Q}_{11}+{\cal Q}_{12}$ and of ${\cal Q}_{13}$ independently and
whose $m^2$ in units of $\frac{g'^2}{\pi a^2}$ go like
$(A+Bg'^2)^2$, where $A,B$ are some constants; (iv) those massless
states which become zero upon the action of ${\cal Q}_{11}+{\cal
Q}_{12}+{\cal Q}_{13}$. From Fig.~\ref{spectrum}(a) and (b) it is
easy to identify one state each for the second and third type
because $m^2$ of a state of the second type go like $g'^4$, giving
rise to a straight line with a non-zero slope for all $g'$ in the
log-log plot, while $m^2$ for the third type is $(A+Bg'^2)^2$,
leading to some flat, constant line at small $g'$ and a (inclined)
straight line at large $g'$. We should note that for the second kind
one should take into account the level crossing. The rest of the
states clearly fall into either the first kind or the fourth kind.
States of the first type yield $g'$-independent $m^2$, thus, a flat
line in the log-log plot, while states of the fourth type are
massless represented by the ``dots" below the line of $\log_{10}
m^2=-8$ since the numerical zero is set to $10^{-8}$ in our code.

\begin{figure}[t!]
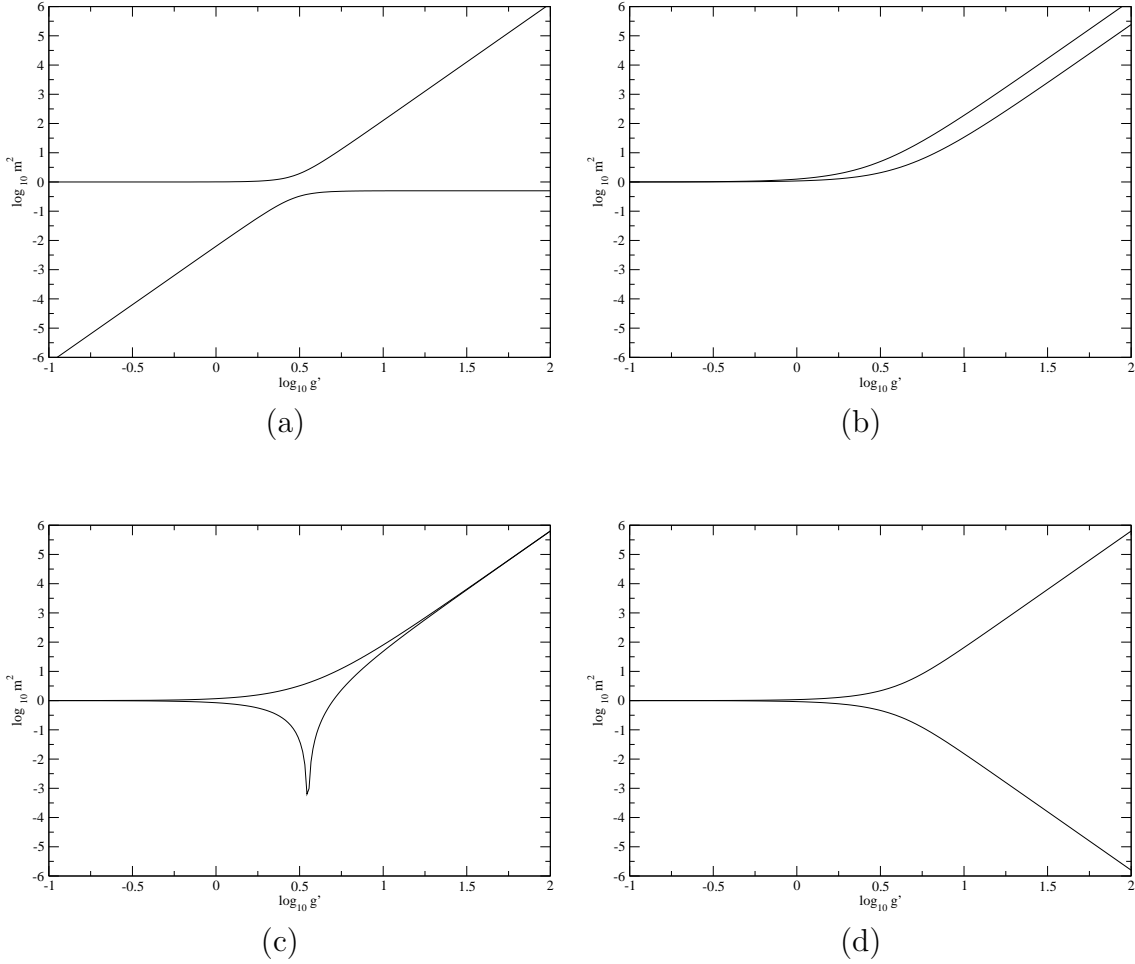

\begin{tabular}{cc}
\epsfig{figure=sampleA.eps,width=7.3cm,angle=0}&
\epsfig{figure=sampleB.eps,width=7.3cm,angle=0}\\
\vspace{1cm}
(a)&(b)\\
\epsfig{figure=sampleD.eps,width=7.3cm,angle=0}&
\epsfig{figure=sampleE.eps,width=7.3cm,angle=0}\\
(c)&(d)
\end{tabular}

\caption{Sample spectra obtained from our toy model. (a) and (b) can
be seen in the actual full spectrum in Fig.~\ref{spectrum}, while
(c) and (d) cannot.} \label{toy:samples}
\end{figure}

This discussion does not however seem to explain the dependence on
$g'$ of the mass spectrum with $K=5,6$. To get the full
understanding of the behavior, we made a toy model. In this model we
have a $2\times 2$ matrix $R$ for the boson sector of $Q^-$ given by
\[ R =\left(\begin{array}{cc} b_1+c_1g'^2&b_2+c_2g'^2 \\
    b_3+c_3g'^2&b_4+c_4g'^2\end{array}
   \right),\]
where $b_i$ with $i=1,2,3,4$ is equal to either 0 or 1 and $c_i$ is
equal to either 0 or $1/4\pi$. Here one should notice that we've
factored out $g'$ from $R$ or $Q^-$, and therefore $g'^2$ from
$P^-$. The $Q^-$ for this toy model is thus given by
\[ Q^-/g' =\left(\begin{array}{cc} 0&R \\ R^T&0\end{array}
   \right),\]
where $T$ stands for the transpose. Thus, the matrix to diagonalize
is
\[ (Q^-/g')^2 =\left(\begin{array}{cc} RR^T&0 \\ 0&R^TR \end{array}
   \right),\]
or equivalently $RR^T$. Among the $2^8=256$ possible forms for
$Q^-$, we found sets of parameters that lead to a level crossing,
and non-trivial behaviors in the mass spectrum. Some of those
non-trivial ones look the same as some of those in
Fig.~\ref{spectrum}, while there are others which do not look like
any of those in Fig.~\ref{spectrum}. For example see
Fig.~\ref{toy:samples}, where Fig.~\ref{toy:samples}(a) and (b) are
the ones that we can see in the actual spectrum in
Fig.~\ref{spectrum}, while \ref{toy:samples}(c) and (d) are not. The
sets of parameters we used are given in Table \ref{parameters}. Of
course there are ones which are seen in Fig.~\ref{spectrum}, but
cannot be found in our toy model. However, it is very likely that as
we increase the size of the matrix $R$ of our toy model, we would be
able to identify those not-yet-seen behaviors in our toy model as
well.

\begin{table}[h]
\begin{center}
\begin{tabular}{|c|cccccccc|}
\hline
&$b_1$&$c_1$&$b_2$&$c_2$&$b_3$&$c_3$&$b_4$&$c_4$\\
\hline
Fig.~\ref{toy:samples}(a)&0& $1/4\pi$& 0& $1/4\pi$& 1& 0& 0& 0\\
\hline
Fig.~\ref{toy:samples}(b)&0& 0& 1& $1/4\pi$& 1& $1/4\pi$& 0& $1/4\pi$\\
\hline
Fig.~\ref{toy:samples}(c)&0& 0& 1& 0& 1& 0& 0& $1/4\pi$\\
\hline
Fig.~\ref{toy:samples}(d)&0& $1/4\pi$& 1& 0& 1& 0& 0& $1/4\pi$\\
\hline
\end{tabular}
\end{center}
\caption{Parameter sets used for our toy model to get each of the
spectra in Fig.~\ref{toy:samples}.} \label{parameters}
\end{table}

Using this toy model, we can study wavefunction dependence on the
coupling $g'$. As the simplest example, consider the case of the
level crossing shown in Fig.~\ref{toy:samples}(a). In this case we
can think of a bound state $|m^2\ket$ as a linear combination of two
different states,
\[ |m^2\ket =f(g')|1\ket+h(g')|2\ket, \]
where $f(g')$ and $h(g')$ are wavefunctions, which depend on $g'$.
$|1\ket$ is a state of the first type of the four we considered
above and responsible for the constant behavior of the mass spectrum
and $|2\ket$ is a state of the second type responsible for the
$g'^4$-behavior. In Fig.~\ref{toy:samples}(a) the higher energy
state stays constant for small $g'$, where $f(g')\gg h(g')$, and
goes like $g'^4$ for large $g'$, where $h(g')\gg f(g')$. The
opposite behavior of the wavefunctions is true for the lower energy
state. That is, the lower energy state goes like $g'^4$ for small
$g'$, where $h(g')\gg f(g')$, and stays flat for large $g'$, with
$f(g')\gg h(g')$. This observation implies that for more general
cases a bound state is a linear combination of states of the four
types associated with $g'$-dependent wavefunctions, and it is the
non-trivial $g'$-dependence of the wavefunctions that gives rise to
such a rich, complicated spectrum in Fig.~\ref{spectrum}.

We expect that the structure of the mass spectrum as a function of
$g'$ will persist for the cyclic states and in fact we have
numerically confirmed the similar structure for them as well.

Note that since the dominant structure of a bound state changes as
one changes $g'$, there is some sort of ``transition" as one goes
from weak coupling to strong coupling. It is of great interest to
see if the winding number dependence of the mass spectrum varies due
to this transition. We are not able to identify any states in strong
coupling regime because of the rich and complicated behavior of the
spectrum although we are able to find some states in the
intermediate region where $g'=1$.

We discuss the mass spectrum of the cyclic states as a function of
the winding number and the resolution with $g'=1$ in more detail in
the next section. The discussion of the mass spectrum of the
non-cyclic states is in the following section.

\section{Numerical results for the cyclic $(W_I\ne 0)$ bound states}
In principle we can study the case where both of the winding numbers
are non-zero and the case where one of them equals zero. However,
the size of the Fock basis is much larger for the former case than
for the latter. This means that we can reach a higher resolution for
the latter case. Thus, in order to get enough data to analyze for
our first attempt we restrict ourselves to the case where we set one
of the winding numbers to zero. Since we have two $Z_2$ symmetries,
$(W_1,W_2)\leftrightarrow (W_2,W_1)$ and $(W_1,W_2)\leftrightarrow
(-W_1,-W_2)$, we can set $W_2=0$ without loss of generality and
consider only positive $W_1$ when studying the winding number
dependence of the bound states.

As guaranteed by the super--algebra, we find numerically a
degeneracy in the mass spectrum between massive fermionic and
bosonic states. However, this supersymmetry is broken for the
massless states since we do not preserve the entire set of super
symmetry algebra. In this section we only consider the massive bound
states, and therefore it suffices to consider only bosonic states.

\begin{figure}[t!]
\begin{tabular}{cc}
\epsfig{figure=CycStateA.eps,width=7.3cm,angle=0}&
\epsfig{figure=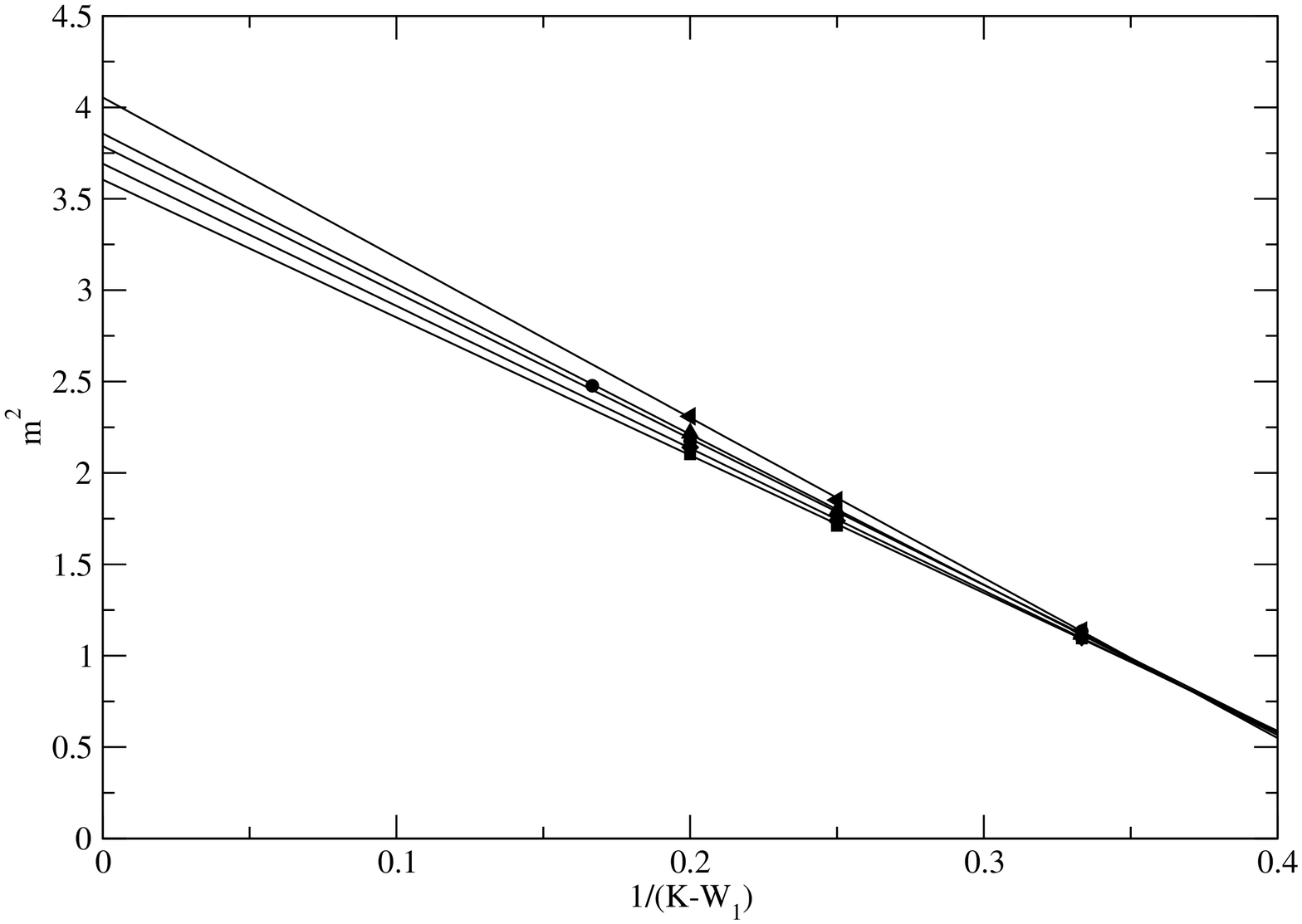,width=7.3cm,angle=0}\\
\vspace{1cm}
(a)&(b)\\

\epsfig{figure=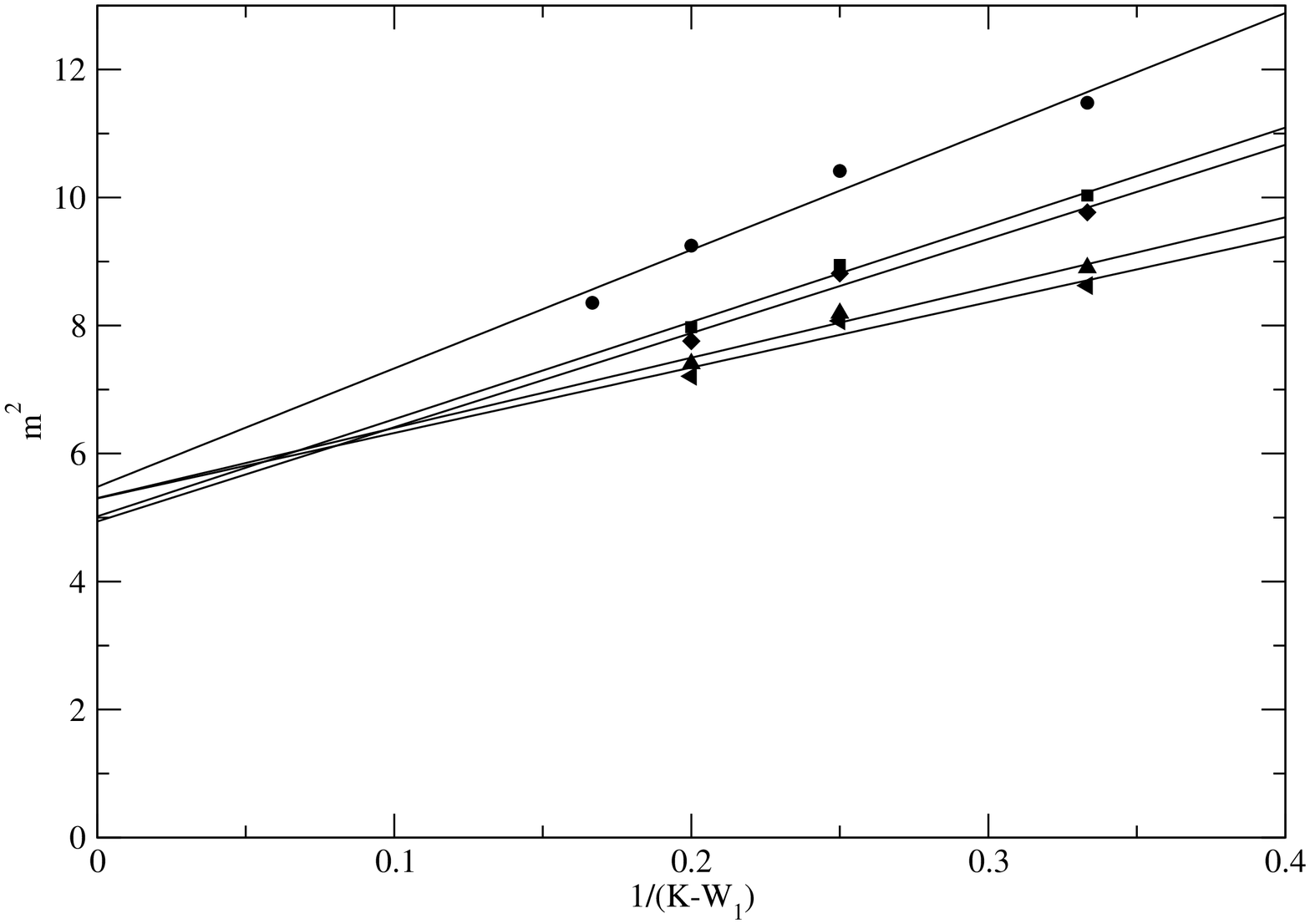,width=7.3cm,angle=0}&
\epsfig{figure=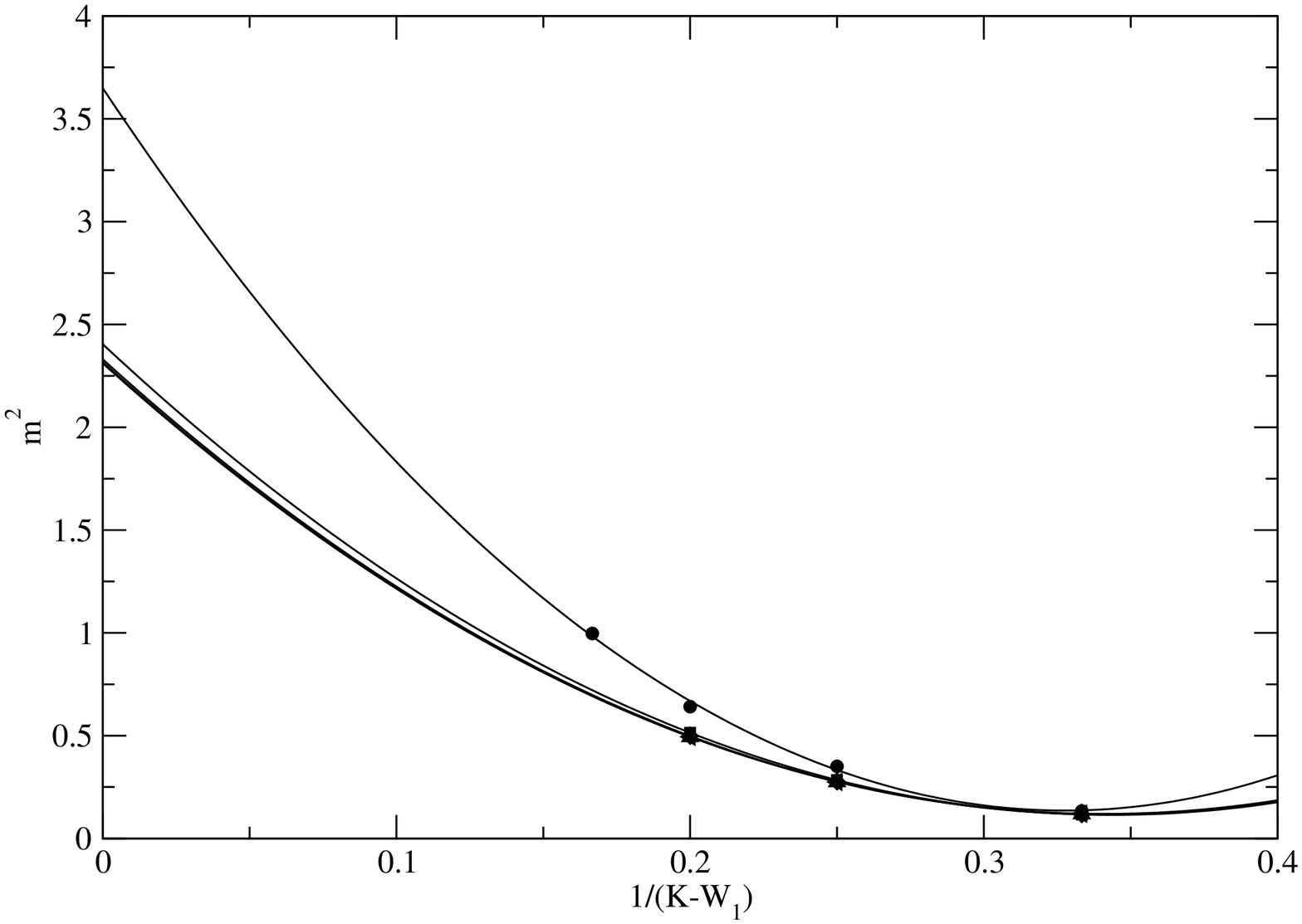,width=7.3cm,angle=0}\\
(c)&(d)

\end{tabular}

\caption{Plots of $m^2$ in units of $\frac{g'^2}{\pi a^2}$ of
low-energy cyclic bound states versus $1/(K-W_1)$ for
$W_1$=1(circle), 2(square), 3(diamond), 4(triangle up), 5(triangle
left). Also shown are a linear fit for (a), (b), and (c) and a
quadratic fit for (d). The coupling $g'\equiv g\sqrt {N_c}=1$.}
\label{fit:cyclic}
\end{figure}

In Fig.~\ref{fit:cyclic}(a), (b), (c), and (d) we give plots of
$m^2$ with $g'=1$ for four low--energy bound states as a function of
$1/(K-W_1)$ and extrapolate $m^2_\infty$ in the $(K-W_1) \to \infty$
limit using a linear fit $b+c/(K-W_1)$ for (a) through (c) and a
quadratic fit $b+c/(K-W_1)+d/(K-W_1)^2$ for (d), where $b,c,d$ are
fitting parameters. We identify a bound state with different $K$'s
from the properties of the bound state, such as the averaged number
of partons of a particular type etc. We present here four bound
states we could easily identify. The dominate fock component of the
bound state in (a) and (c) has the form $b^{1\dag}a^{1\dag}\cdots
a^{1\dag}b^{1\dag}$. For the bound state in (b) the dominant
component is of the form $b^{1\dag}a^{1\dag}\cdots a^{1\dag}
b^{2\dag}$. The bound state in (d) has the dominant component of
$d^{2\dag}a^{1\dag}\cdots a^{1\dag} a^{2\dag}$.

\begin{figure}[!ht]
\begin{tabular}{cc}
\epsfig{figure=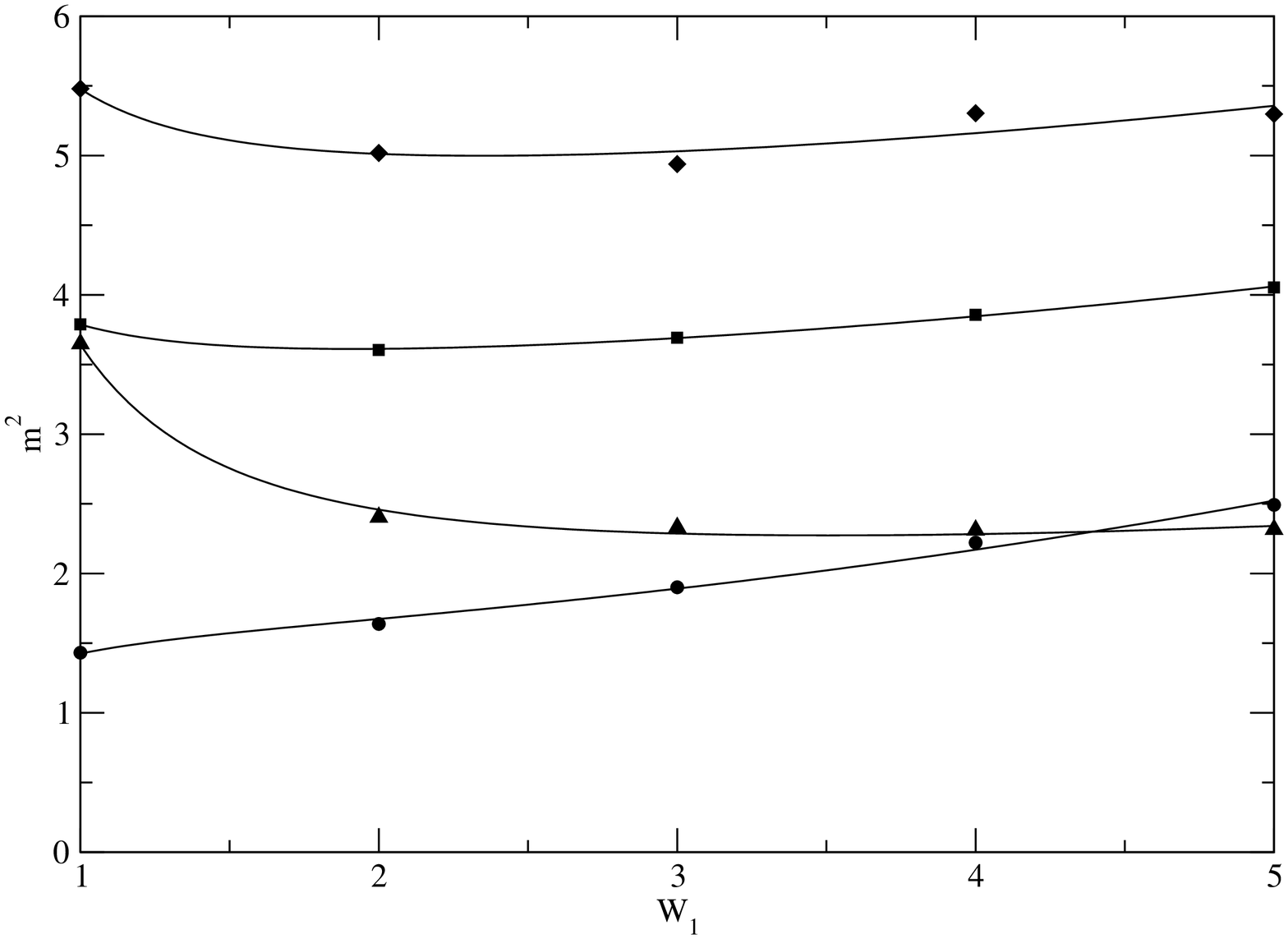,width=7.2cm,angle=0}& \
\epsfig{figure=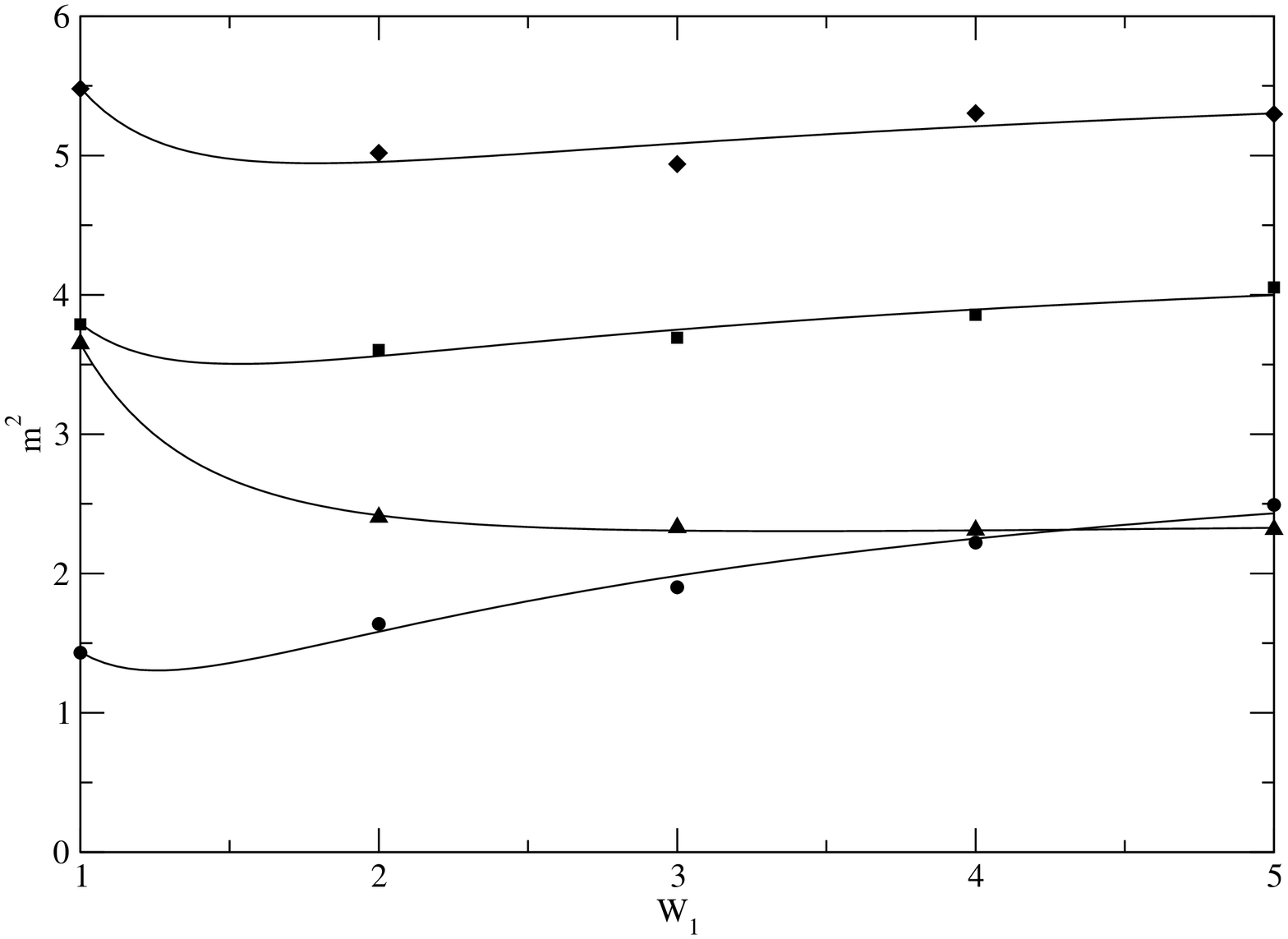,width=7.2cm,angle=0}\\
(a)&(b)
\end{tabular}

\caption{Plots of $K \to \infty$ limit of $m^2$ in units of
$\frac{g'^2}{\pi a^2}$ of low energy cyclic bound states versus
$W_1$ with a fit to the data of the form $b+cW_1^2+d/W_1^2$ in (a)
and of the form $b+c/W_1+d/W_1^2$ in (b). The circles correspond to
the bound state in Fig.~\ref{fit:cyclic}(a), squares in
\ref{fit:cyclic}(b), diamonds in \ref{fit:cyclic}(c), and triangles
in \ref{fit:cyclic}(d).} \label{fit:cyclic2}
\end{figure}

In Fig.~\ref{fit:cyclic2} we present $m^2_\infty$, obtained in
Fig.~\ref{fit:cyclic}(a), (b), (c), and (d), as a function of $W_1$.
 We show a fit to the data of the form $b+cW_1^2+d/W_1^2$ in
Fig.~\ref{fit:cyclic2}(a) and of the form $b+c/W_1+d/W_1^2$ in
Fig.~\ref{fit:cyclic2}(b). As can be seen, it is difficult to say
which fit is better from the graphs. The fit of the form
$b+cW_1^2+d/W_1^2$ appears a bit better.

The use of a fit of the form $b+cW_1^2+d/W_1^2$ has a string theory
justification. In the string theory the energy of a string confined
in one dimension with a period $L$ is given by the sum of its
momentum mode and its winding mode, so that $E=p2\pi/L+qTL$, where
$p,q$ are integers and $T$ is the string tension. Now if we consider
our cyclic bound states as a string confined in the $x_1$-direction
with $L=aW_1$, then it follows that $m^2=b+cW_1^2+d/W_1^2$.

We should however remind the reader that we used a fit of the form
$b+c/W+d/W^2$ in Ref.~\cite{Harada:2003bs}. There we argued that the
operator has the form $Q^-=b+ck_{\perp}$ in 2+1 continuous theory
and $m^2\sim (Q^-)^2=b+c/W+d/W^2$ with $k_{\perp}\sim 1/L\sim 1/W$.
This behavior is consistent with the unique properties of SYM
theories that we have seen in previous SDLCQ calculations
\cite{Antonuccio:1998kz}. We have seen that as we increase $K$ we
uncover longer bound states that have lower masses. Supersymmetric
theories like to have light bound states with long strings of
gluons. We call these bound states with long strings of gluons,
stringy bound states. In 3+1 dimensions with two transverse lattices
we have seen the stringy bound states as well, and we have
$Q^-=b+ck_{1}+dk_2$, leading to the fit of the form
$b+c/W_1+d/W_1^2$ in Fig.~\ref{fit:cyclic2}(b) for $k_1\sim 1/L\sim
1/W_1$ and $k_2=0$. Up to the numerical resolution we can correctly
reach, we can not say for sure which form of $m^2$ describes ${\cal
N}$=(2,2) SYM in 3+1 dimensions. It appears that the form
$b+cW_1^2+d/W_1^2$ is preferable, suggesting that the cyclic bound
states in 3+1 dimensions are more like a string with the energy of
the form $E=p2\pi/L+qTL$.

\section{Numerical results for the non-cyclic bound states $(W_I=0)$}

\begin{figure}[!ht]
\begin{center}
\epsfig{figure=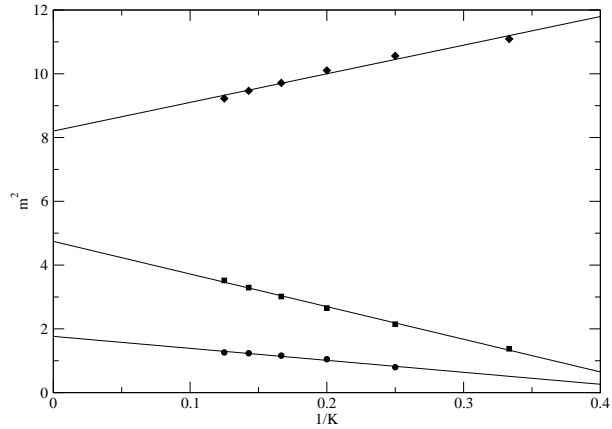,width=8cm}

\end{center}

\caption{Plots of $m^2$ in units of $\frac{g'^2}{\pi a^2}$ of
low--energy non-cyclic bound states against $1/K$ with a linear fit
to the data. The coupling $g'\equiv g\sqrt {N_c}=1$. The circles
correspond to bound state A, squares to the state B, diamonds to
state C} \label{states}
\end{figure}

Let us now discuss numerical results for the non-cyclic bound
states. Again we follow bound states that we can easily identify
from the properties of the bound states. In Fig.~\ref{states} we
show $m^2$ of three low-energy states in units of $\frac{g'^2}{\pi
a^2}$ as a function of $1/K$ with $g'\equiv g\sqrt{N_c}=1$. The
state A denoted by circles is composed primarily of two bosons and
two fermions, $b^{1\dag}d^{1\dag}a^{1\dag}b^{1\dag}$. The state B
and C denoted by squares and diamonds are composed primarily of two
fermions, $b^{1\dag}b^{2\dag}$ and $b^{1\dag}b^{1\dag}$,
respectively. We show a linear fit to the data and see good
conversion as $K\to \infty$ for all the three states. The
extrapolated values for $m^2$ in the limit of $K\to \infty$ are
given in Table~\ref{masses}. We also find the stringy states for the
non-cyclic states.

\begin{table}
\begin{center}
\begin{tabular}{|c|c|c|c|}
\hline
&$b^{1\dag}d^{1\dag}a^{1\dag}b^{1\dag}$&$b^{1\dag}b^{2\dag}$& $b^{1\dag}b^{1\dag}$ \\
\hline $m^2_{\infty}$&1.764 &4.744 &8.204\\
\hline
\end{tabular}
\end{center}
\caption{Extrapolated values for $m^2$ in units of $\frac{g'^2}{\pi
a^2}$ as $K \to \infty$ for State A, B, and C in Fig.~\ref{states}
represented by its dominant Fock state.} \label{masses}
\end{table}

Recall that we found in Sec.~4 that a bound state would be a linear
combination of states of the four types we enumerated in Sec.~4.
Hence, it is instructive to see if we can identify the three bound
states with any of the four types. For $K=4$ we can identify all the
three bound states with those that are killed by ${\cal Q}_{13}$ and
whose mass in units of $\frac{g'^2}{\pi a^2}$ are independent of
$g'$. However, as $K$ increases, we are not able to classify them
into any particular type of the four. This is because as we increase
$K$ the number of states becomes very large and the mass spectrum
becomes dense. It is likely that these states mix with other nearby
states with the same coupling dependence, giving rise to small
changes in $m^2$ but still the same general coupling dependence. At
this time however we are not able to resolve the spectrum in an
enough detail to study these effects.

\section{Discussion}

We have presented the standard formulation of ${\cal N}$=1 SYM in
3+1 dimensions with a two spatial dimensional transverse lattice.
Then we gave the SDLCQ formulation of the theory. We found that the
standard formulation suffers from a fermion species doubling
problem, while SDLCQ formulation does not. In the frame where the
transverse momenta equal to zero, ${\cal N}$=1 SUSY in 3+1 dimensions
is equivalent to ${\cal N}$=2 SUSY in 1+1 dimensions also known as
${\cal N}$=(2,2) SUSY. We were able to present $Q^-_{\a}$ which has
the correct continuum form and yields by the SUSY algebra a discrete
form of $P^-$, where $\a=1,2$. This $P^-$ then coincides with its continuum
form in the continuum limit. Since $Q^-_1$ and $Q^-_2$ don't commute with
each other in our formulation, we are to use only one of them to
solve the mass eigenvalue problem, preserving one exact SUSY.

We found that this $Q^-_{\a}$ consists of terms which are
proportional to $g'\equiv g\sqrt{N_c}$ and terms which go like
$g'^3$. This led us to investigate in some detail the $g'$
dependence of the mass spectrum. From a simple toy model we
concluded that the rich, complicated behavior of the mass spectrum
with varying $g'$ is due to some non-trivial coupling dependence of
the wavefunctions. This is also responsible for a ``transition" in
the structure of a bound state when going from weak coupling to
strong coupling. Because the dominant structure of a bound state
changes with changing $g'$.

We classified the bound states into two types, the cyclic and
non-cyclic as we did in Ref.~\cite{Harada:2003bs}. The cyclic bound
states are those whose color flux goes all the way around in one or
two of the transverse directions. The bound states whose color flux
is localized and does not wind around are referred to as the
non-cyclic bound states. For each type of the bound states, we were
able to identify some bound states in the mass spectrum for $g'=1$
and found the $K\to\infty$ limit of $m^2$.

For the cyclic bound states we were able to present $m^2$ as a
function of the winding number $W_I$ in the $x_I$ direction with
$I=1,2$. We found two very good fits to the data. The first fit
$b+cW_I^2+d/W_I^2$ is motivated by the string theory, where the
energy has the form $E=p2\pi/L+qTL$, where $p,q$ are some integers,
$T$ is the string tension and $L$ is the period of the transverse
lattice. The other fit $b+c/W_I+d/W_I^2$ is motivated by the
operator structure of $Q^-_{\a}$. It appeared that $b+cW_I^2+d/W_I^2$ is
preferable,

For the non-cyclic states as $1/K\to 0$ we saw good linear
conversion of $m^2$ of the low-energy bound states that we could
identify and gave the extrapolated values for $m^2$. We could
identify for $K=4$ the bound states with a state whose $m^2$ in
units of $\frac{g'^2}{\pi a^2}$ are independent of $g'$ though we
were not able to do so for higher $K$'s because of the dense, and
complicated spectrum.

In summary, we were able to present a formulation of SYM in 3+1
dimensions with one exact SUSY on a two dimensional transverse
lattice and find the mass spectrum nonperturbatively. There remain
however a number of important questions to answer. First and
foremost it is of great importance to determine the form of $m^2$
numerically to better precision. It is interesting to see what the
winding number dependence of $m^2$ is if both of the winding numbers
are non-zero. We need to invent a method to resolve the dense
spectrum at strong couplings. This will help us see if there is any
``transition" in the form of $m^2$ as one goes from weak coupling to
strong coupling. However, perhaps most importantly, as discussed in
appendix \ref{reduction} we need to know to what extent we've
resolved the problem caused by the linearization of the link
variables that we needed to quantize the fields. Knowing this tells
us how reliable our numerical results are. Because one of our major
simplifications in numerical calculation in the large $N_c$ limit
comes about from the reduction of the transverse degrees of freedom
whose justification relies upon the presence of the quantized fields
and the vacuum. Restoration of SUSY for massive bound states, which
has been broken by the linearlization gives us some confidence that
our formulation indeed provides some sensible results. However, we
would still have to clarify the issue to be more certain and
confident. To this end, we need to compare our numerical results
with some well-established theoretical predictions and with other
numerical results obtained from the usual lattice calculation.
Hence, it is of importance to apply our formulation to some other
supersymmetric theories in higher than 1+1 dimensions, for instance,
Wess-Zumino model, lattice sigma model, and SQED. It appears that
the application is relatively straightforward. From more practical
point of view, a next question to ask is what happens if one
includes scalars and their superpartners in theory. We did not
consider this case in this paper simply because this was the first
attempt to formulate SYM in 3+1 dimensions with one exact SUSY on a
two dimensional transverse lattice and, thus, we wanted to consider
the simplest possible case. However, it is of great interest to
consider the question in the future. The authors believe that when
we are able to answer all those questions, we will also be able to
test the predictions made by Armoni, Shifman and Veneziano
\cite{Armoni:2003gp, Armoni:2003fb}.

\section*{Acknowledgments}
This work was supported in part by the U.S. Department of Energy.

\appendix

\section{Eguchi-Kawai Reduction} \label{reduction}
For our numerical calculation we've set $N_{sites}=1$, in other
words, we've dropped the site indices. This reduction of the
transverse degrees of freedom has brought a great amount of
simplification in our calculation and needs some detailed
justification. Since it is only the supercharges that we need to do
our calculation, if the supercharges do not depend on the site
indices in the large $N_c$ limit, neither does any quantity that can
be computed from $Q^-_{\a}$, for instance $m^2$ for our case.
Therefore, in order to justify the reduction of the degrees of
freedom for our purposes, it suffices to show the independence of
$Q^-_{\a}$ of the site indices in the large $N_c$ limit. In this
appendix, in particular, we will show that in the large $N_c$ limit
the leading order terms of the supercharges $Q^-_{\a}$ with keeping
all the site indices are the same as those with setting
$N_{sites}=1$. We should note that this sort of arguments about the
justification of the reduction on a transverse lattice have already
been given in literature, for instance see
Refs.~\cite{rev,Dalley:1998bj,Dalley:1997an} and our arguments below
closely parallel those in the Refs.~\cite{rev,Dalley:1997an}.

In what follows we only consider $Q^-_1$, however the same arguments
apply equally well to $Q^-_2$. For definiteness let us first
consider a Fock state denoted by
\[ \sum _n \tr[\ldots d^{1\dag}_n(k_1)a^{1\dag}_n(k_2)
b^{1\dag}_n(k_3)a^{1\dag}_{n-i_1}(k_4) \ldots ]|0\ket,\] where we've
written $k^+\equiv k$, $n\equiv (i,j)$ is the transverse lattice
site, $i_1$ is the vector of length $a$ pointing the $x^1$
direction, $a$ is the lattice spacing, and the dots represent some
creation operators. When we act on this state with $Q^-_1$, we get
for example from one of the terms in $Q^-_1$, say
$b^{2\dag}d^{1}a^{1}\equiv \sum_n
\tr[b^{2\dag}_n(p_1+p_2)d^{1}_n(p_1)a^{1}_n(p_2)]$ on it
\[  N_c\sum_n\tr[\ldots b^{2\dag}_n(k_1+k_2)b^{1\dag}_n(k_3)
a^{1\dag}_{n-i_1}(k_4)\ldots ]|0\ket. \] If we set $N_{sites}=1$,
then the Fock state now becomes
\[  \tr[\ldots d^{1\dag}(k_1)a^{1\dag}(k_2)b^{1\dag}(k_3)
a^{1\dag}(k_4)\ldots ]|0\ket,\] and upon the action of $Q^-_1$ we
get from $b^{2\dag}d^{1}a^{1}\equiv
\tr[b^{2\dag}(p_1+p_2)d^{1}(p_1)a^{1}(p_2)]$ on it
\begin{equation}
N_c\tr[\ldots b^{2\dag}(k_1+k_2)b^{1\dag}(k_3)a^{1\dag}(k_4) \ldots
]|0\ket, \label{leading}
\end{equation}
{\it and} one more term
\begin{equation}
\tr[\ldots b^{2\dag}(k_1+k_4)\ldots
]\tr[a^{1\dag}(k_2)b^{1\dag}(k_3)]|0\ket. \label{extra}
\end{equation}

Notice that the extra term Eq.~\eqref{extra} we get by setting
$N_{sites}=1$ is down by $1/N_c$ compared to the leading order term
Eq.~\eqref{leading} and thus we can ignore it in the large $N_c$
limit. Of course, in the above example, we could and would have
gotten many more terms depending on what we have in those 'dots'
inside the trace of the Fock state we considered. However, it is
easy to see that our conclusion remains the same; all the extra
terms we get by having only one site are down by $1/N_c$ or more
powers of $1/N_c$. This all comes down to the fact that we can have
only single-traced states in the large $N_c$ limit. Therefore, we
find that the leading order terms of $Q^-_{\a}$ are the same whether
we keep track of the site indices or not. Although this proof is for
finite $K$, we suspect that the same result would persist at
infinite $K$.

The way to justify the reduction here should be contrasted to the
way exploited by Eguchi and Kawai \cite{egk83}. Eguchi and Kawai
showed that in the large $N_c$ limit we can work with only one
lattice site in each of the space-time directions in Euclidean
space. However, the proof was based on, among others, the assumption
that $U(1)^d$ symmetry is not spontaneously broken, where $d$ is the
number of the space-time dimensions. This assumption was found to be
wrong for $d>2$ at weak couplings by the authors of
Ref.\cite{Bhanot:1982sh}. To resolve this problem, there have been
many models proposed, for instance quenching \cite{Bhanot:1982sh}
and twisted \cite{Gonzalez-Arroyo:1982hz} lattice formulations. Here
in our formulation, however, we believe that we do not have to
introduce any of the modified lattice formulation since the way we
justify the reduction is quite different the way Eguchi and Kawai
do. Our proof stands on its own feet regardless of us maintaining
the $U(1)^d$ symmetry or not and, therefore, would not suffer from
the problem associated with the naive Eguchi-Kawai reduction as we
go from weak to strong couplings.

A question, however, remains. That is the question of how well we've
managed to quantize the fields since all our arguments above rely on
the fact that we have the quantized fields and true vacuum. Put in
another way, how good the reduction procedure is depends on how good
our quantization procedure is. Recall that to quantize, we had to
``linearize'' the unitary link variables, which leads to the
breakdown of SUSY. The authors of
Refs.~\cite{rev,Dalley:1998bj,Dalley:1997an} make use of the
``color-dielectric'' formulation to resolve the problem for
non-supersymmetric theories. Although this formulation resolves the
problem completely, it prevents one from going to small lattice
spacings. In our formulation we do not have that constraint on the
lattice spacing. However, the price we pay is that we resolve the
problem of the linearization partially, not completely. Thus, it is
of great importance for one to see to what extent we've resolved the
problem and, if possible and necessary, to find a way to get around
it completely. Up to this point we are not able to answer this
question, but this is one of the crucial steps we should take
towards a more sensible supersymmetric model on a lattice within our
formulation.

\section{${\cal N}$=1 super-algebra in Majorana representation}
\label{majorana}

In this appendix we give the super-algebra in Majorana
representation in $D+1$ dimensional light-cone coordinates where
$D=1,2,3$.

In Majorana representation Majorana spinors have real component
fields, and can be written as
\[
   \Psi_M=\left(\begin{array}{c}\theta_L \\ \theta_R\end{array}\right),
\]
where $\theta_L$, $\theta_R$ are left-moving, right-moving spinors
with real components. This implies that the supercharge $Q$ is also
a Majorana spinor with real components of the form
\[
   Q=\int d^{D}x j^+=\left(\begin{array}{c}Q_L \\ Q_R\end{array}\right)
    \equiv \left(\begin{array}{c}Q^+ \\ Q^-\end{array}\right),
\]
where the integration is taken over the $D$ spatial dimensions,
$j^{\mu}$ is the supercurrent, which is a Majorana spinor.

In terms of the Majorana super-charge, the super-algebra is given by
\begin{equation}
\{Q,\bar Q\}=2\Gamma^{\mu}P_{\mu}, \label{susymajorana}
\end{equation}
where $\bar Q\equiv Q^{\dag}\Gamma^0$ in any representation, and
thus $\bar Q =Q^T\Gamma^0$ in Majorana representation.

\subsection{D=1}
For 1+1 dimensional case, we have $\Gamma^0=\sigma^2$ and
$\Gamma^1=i\sigma^1$, so that
\[
    \Gamma^+\equiv \frac{\Gamma^0+\Gamma^1}{\sqrt 2}
    =i\left(\begin{array}{cc} 0& 0\\
            \sqrt 2&0\end{array}\right), \qquad
    \Gamma^-\equiv \frac{\Gamma^0-\Gamma^1}{\sqrt 2}
    =i\left(\begin{array}{cc} 0& -\sqrt 2\\
            0&0\end{array}\right)\]
and $\bar Q=i(Q^-,-Q^+)$. Thus, Eq.~\eqref{susymajorana} reads
\[ \{Q,\bar Q\}=2\Gamma^{\mu}P_{\mu}=i\left(\begin{array}{cc}
    0& -2\sqrt 2P_-\\ 2\sqrt 2 P_+ &0\end{array}\right)
    =i\left(\begin{array}{cc}
    0& -2\sqrt 2P^+\\ 2\sqrt 2 P^- &0\end{array}\right), \]
or
\[ \{Q^{\pm},Q^{\pm}\}=2\sqrt 2 P^{\pm}, \quad
    \{Q^{+},Q^{-}\}=0. \]

\subsection{D=2}
In this case $\Gamma^0=\sigma^2$, $\Gamma^1=i\sigma^1$ and
$\Gamma^2=\Gamma^{\perp}=i\sigma^3$. Therefore,
\[ \{Q,\bar Q\}=2\Gamma^{\mu}P_{\mu}=i\left(\begin{array}{cc}
    2P_{\perp}& -2\sqrt 2P_-\\ 2\sqrt 2 P_+ &-2P_{\perp}
    \end{array}\right)
    =i\left(\begin{array}{cc}
    -2P^{\perp}& -2\sqrt 2P^+\\ 2\sqrt 2 P^- &2P^{\perp}
    \end{array}\right), \]
or
\[ \{Q^{\pm},Q^{\pm}\}=2\sqrt 2 P^{\pm}, \quad
    \{Q^{+},Q^{-}\}=-2P^{\perp}   . \]

\subsection{D=3}
In 3+1 dimensions Majorana spinors have four components and thus the
supercharge can be written as
\[ Q=\left(\begin{array}{c} Q^+ \\Q^-\end{array}\right)\equiv
    \left(\begin{array}{c} Q^+_1\\Q^+_2 \\Q^-_1\\Q^-_2\end{array}\right),
    \quad \bar Q=Q^T\Gamma^0=i(Q^-_2,-Q^-_1,Q^+_2,-Q^+_1).\]
Gamma matrices are $4\times 4$ matrices given by
\[ \Gamma^0\equiv \left(\begin{array}{cc} 0&\sigma_2 \\ \sigma_2&0\end{array}
   \right), \quad
   \Gamma^1\equiv i\left(\begin{array}{cc} \sigma_1&0 \\ 0&\sigma_1\end{array}
   \right),\quad
   \Gamma^2\equiv i\left(\begin{array}{cc} \sigma_3&0 \\ 0&\sigma_3\end{array}
   \right),\quad
   \Gamma^3\equiv \left(\begin{array}{cc} 0&-\sigma_2 \\ \sigma_2&0\end{array}
   \right),
\]
\[  \Gamma^{+}\equiv\frac {\Gamma^0+\Gamma^3}{\sqrt 2}=\left(\begin{array}{cc}
    0&0 \\\sqrt 2\sigma_2& 0\end{array}\right),\quad
    \Gamma^{-}\equiv\frac {\Gamma^0-\Gamma^3}{\sqrt 2}=\left(\begin{array}{cc}
    0&\sqrt 2\sigma_2 \\ 0& 0\end{array}\right).
\]
Then Eq.~\eqref{susymajorana} yields
\begin{eqnarray*}
 && \{Q,\bar Q\}=2\Gamma^{\mu}P_{\mu}=2\left(\begin{array}{cc}
    i\sigma^1P_1+i\sigma^3P_2&\sqrt 2\sigma^2P_-\\
    \sqrt 2\sigma^2P_+ &i\sigma^1P_1+i\sigma^3P_2
    \end{array}\right)\\
 && =2i\left(\begin{array}{cccc}
    -P^2&-P^1&0&-\sqrt 2 P^+\\-P^1& P^2&\sqrt 2 P^+&0\\
    0&-\sqrt 2P^-&-P^2&-P^1 \\ \sqrt 2P^-&0&-P^1&P^2
    \end{array}\right).
\end{eqnarray*}
Hence, we find
\[
  \{Q^{\pm}_{\a},Q^{\pm}_{\beta}\}
    =2\sqrt 2 P^{\pm}\delta_{\a\beta},
\]
\[
    \{Q^{+}_{1},Q^{-}_{1}\}=-\{Q^{+}_{2},Q^{-}_{2}\}=2P^1, \quad
    \{Q^{+}_{1},Q^{-}_{2}\}=\{Q^{+}_{2},Q^{-}_{1}\}=-2P^2,
\]
where $\a,\beta=1,2$. Note that if $P^1=P^2=0$, then this algebra
coincides with the one for ${\cal N}$=2 SUSY in 1+1 dimensions also
known as ${\cal N}$=(2,2) SUSY.

\end{document}